\documentclass[11pt,a4paper]{article}
\usepackage{jinstpub}
\usepackage{lineno}
\usepackage{algorithm}
\usepackage{algorithmic}
\usepackage{float}  
\usepackage{graphicx}

\usepackage{amsmath}
\usepackage{esint}
\usepackage[version=4]{mhchem}
\usepackage{lineno}
\usepackage{subfig}
\usepackage{booktabs}
\usepackage{listings}
\usepackage{xcolor}
\usepackage{url}
\newcommand{\changed}[1]{{\color{blue}#1}}

\lstdefinestyle{jinstcode}{
  language=C++,
  basicstyle=\ttfamily\footnotesize,
  keywordstyle=\bfseries,
  commentstyle=\itshape,
  stringstyle=,
  numbers=left,
  numberstyle=\tiny,
  stepnumber=1,
  numbersep=5pt,
  frame=single,
  breaklines=true,
  columns=fullflexible,
  showstringspaces=false,
  tabsize=2
}

\makeatletter

\pdfoutput=1 

\author[a,1]{Tanay Dey,\note{Corresponding author.}}
\author[a,b]{Suraj Shaw,}
\author[a]{Ritabrata Banerjee,}
\author[a,b]{Pratik Majumdar,}
\author[a,b]{Satyaki Bhattacharya}

\affiliation[a]{Homi Bhabha National Institute,\\Mumbai, India}
\affiliation[b]{Saha Institute of Nuclear Physics,\\Kolkata,India}

\emailAdd{tanay.dey@cern.ch, tanay.jop@gmail.com}

\title{PySiPMGUI: A Universal Python-Based Software for Photodetector I-V Quality Assurance: From Underground Dark Matter Searches to Astroparticle Cherenkov Cameras}

\abstract{Silicon photomultipliers (SiPMs) are currently the most prevalent photon detection technology in modern experiments in high-energy physics, astroparticle physics, neutrino physics, and dark matter searches. The high detection efficiency for photons, excellent timing resolution, small size, and magnetic field independence make them ideal for precision measurements in low-light conditions. However, key parameters like breakdown voltage, gain, and dark count rate show a strong dependence on the bias voltage and temperature, requiring a systematic characterization. 
In this work, we present an open-source graphical user interface (GUI) for automated SiPM characterization, \changed{leveraging standard laboratory instrument communication protocols (PyVISA)}. The tool provides a free, open, and platform-independent solution for detector R\&D and large-scale SiPM characterization, and is available at this GitHuB repository \cite{dey_pysipmgui_2026}. \changed{The analysis procedure is validated against the manufacturer's datasheet for the SensL MicroFC-60035-SMT, yielding results in good agreement for both breakdown voltage and dark count rate.}}


\usepackage{subfig}
\begin{document}
\maketitle
\section{Introduction}
\label{sec:intro}

Silicon Photomultipliers (SiPMs) technology is currently the leading photon-detection technology and widely used in contemporary experiments in numerous fields. 
Examples include detectors used in High Energy Physics (e.g. in plastic scintillator detectors in CMS calorimetry ~\cite{CMS_HCAL_Phase1_TDR, CMS_HCAL_SiPMs_NIMA_2007}), Astroparticle physics (the MAGIC telescope~\cite{MAGIC_Camera_Upgrade}, \changed{FACT}~\cite{BRAUN2009400,NEISE201717}), Neutrino Physics (like the DUNE experiment~\cite{DUNE_TDR}), \changed{and} Dark Matter search experiments (DarkSide-20k experiment~\cite{DarkSide20k_CDR}).

SiPMs offer significant advantages over more conventional photodetectors such as photomultiplier tube (PMT) and avalanche photodiode (APD). They run at low bias voltages, are compact and mechanically robust, insensitive to magnetic fields, and provide high gain that is comparable to PMTs. Moreover, SiPMs deliver superior timing resolution, single-photon sensitivity, higher photon detection efficiency (PDE), low power consumption and easy integration to state-of-the-art readout electronic circuits.

The new generation SiPMs can achieve a PDE of about 40-60 \% in the visible wavelength region ~\cite{Otte_SiPM_Review_2017,Hamamatsu_MPPC_Guide_Section3_2018,Hamamatsu_MPPC_S14160_6050HS,Okumura_SiPM_PDE_UpTo60_2023,onsemi_sipm_datasheet_2022}. On the other hand, a standard photomultiplier tube will have a quantum efficiency of about 20-30 \% in the same spectral range ~\cite{PMT_Hamamatsu_Handbook}. The improved PDE makes SiPMs ideal for low-light and precision photon-counting. These experimental domains pursue very different goals, whether it is to monitor TeV-scale collisions or to hunt for rare low-energy recoil events. Nevertheless, characterization of SiPM is necessary to ascertain the breakdown voltage, gain, dark count rate, photon detection efficiency, \changed{and related quantities}. \changed{Bias voltage and temperature strongly influence the gain, dark count rate, and photon detection efficiency through the overvoltage $V_\text{OV} = V_\text{bias} - V_\text{BD}$, while the breakdown voltage $V_\text{BD}$ is a device parameter that depends primarily on temperature.} Proper characterization leads to maximum performance, accurate and reliable measurements, and comparison with manufacturer data. These kinds of characterization tests are standard \cite{KLANNER201936} and are essential in detector development projects. The characterization and performance of SiPMs used in CMS Hadron Calorimeter (HCAL) front-end electronics can be found in~\cite{CMS_HCAL_SiPMs_NIMA_2007}. In 2010, CALICE built large SiPM-scintillator tile arrays and performed detailed characterization studies; details can be found in~\cite{CALICE_SiPM_TileHCAL_2009}. Recently, the LHCb Scintillating Fibre Tracker has provided significant real-world experience with large SiPM arrays operating under high-rate conditions, including the full calibration and monitoring setup required during Run~3~\cite{LHCb_SciFi_Run3_2025}. In the area of astroparticle physics, similar characterization has been performed for SiPM cameras in Cherenkov telescopes, where the SiPMs are \changed{characterized} based on photon detection efficiency, cross-talk, and temperature effects on gain~\cite{CTA_SiPM_Evaluation_SST_2018,ANDERHUB2011107,OTTE2017106}.

A dedicated GUI proves to be very helpful when a large number of sensors must be tested. A unified, automated interface reduces operator-dependent variability by enforcing standardized measurement sequences (e.g., bias ramps, temperature set-points, waveform triggers) and provides immediate visual feedback to identify issues such as unstable bias supply, mis-triggering, light leaks, or abnormal dark-noise behavior before an entire batch is measured. In addition, integrated metadata capture (device ID, test conditions, calibration constants) and automated file naming and organization are crucial for traceability and subsequent cross-comparisons. To perform these tasks, commercially available software packages such as LabVIEW \cite{NI_LabVIEW} and CAEN HERA \cite{CAEN_HERA} are often used. However, these software solutions are platform-dependent, and are not freely available or easily customizable.

In this paper, we present an open-source Graphical User Interface (GUI) developed in Python, designed to address the characterization needs of the broader particle physics community. The PyVISA library~\cite{PyVISA} has been used to communicate with high-voltage source meters (e.g., Keithley~\cite{Keithley_2450}). This enables \changed{the automated acquisition of DC current measurements simultaneously with the applied voltage. The measured current comprises both the pre-breakdown surface leakage and the post-breakdown avalanche current above $V_{\text{BD}}$\cite{ACERBI201916}.} Such automation is essential for online plotting of the I-V curve and for determining the breakdown voltage ($\changed{V_\text{BD}}$), gain uniformity, and dark count rate (DCR).

At Saha Institute of Nuclear Physics (SINP), Kolkata, the tool is currently used to characterize SiPMs that will be coupled to plastic scintillator modules for a position-sensitive portable muon veto system. This portable system will be used to map the background radiation environment at the Jaduguda Underground Science Laboratory (JUSL) for upcoming dark matter search experiments. Furthermore, the framework will also be utilized for the mass characterization of SiPMs used in Cherenkov telescope cameras in astroparticle physics.

By providing a free of cost, transparent, and platform-independent alternative to commercial software suites, this tool aims to streamline detector R\&D and ensure reproducibility across the global physics community.

\changed{The remainder of this paper is organized as follows. Section~\ref{sec:functionality} describes the software functionality and architecture of the GUI, including the instrument control framework, the safe control and environmental monitoring system, acquisition and analysis modes, and fault detection and data management. Section~\ref{sec:theory} presents the theoretical modeling of the SiPM I-V characteristics, covering the pre-breakdown leakage current, the post-breakdown avalanche current, and their temperature dependence. Section~\ref{sec:setup} details the experimental methodology, including the sensor specifications, the biasing and readout circuitry, and the multi-channel SiPM bias distribution board used for the measurements. Section~\ref{sec:iv_logic} outlines the automated I-V characterization logic implemented for both single-channel and multi-channel operation. Section~\ref{sec:results} presents the results and validation of the software, including the determination of the breakdown voltage and quenching resistance, and compares them against expected/manufacturer values. Finally, Section~\ref{sec:summary} summarizes the key findings and discusses future directions for the development of the software.}

\changed{\section{Software Functionality and Architecture}
\label{sec:functionality}

The control application provides an integrated environment for automated current--voltage (I--V) characterization of SiPM devices, combining instrument control, multi-channel automation, on-line physics analysis, and data management within a single graphical interface. This section first describes the overall software architecture and then details the principal functionalities implemented within it.

\subsection{Software Architecture and Instrument Control}
\label{sec:architecture}
\label{subsec:instrument-control}

The application is written in Python~3 using the \texttt{Tkinter} toolkit for the GUI. Instrument communication is handled through \texttt{PyVISA} and \texttt{pySerial}, and analysis and plotting are done with \texttt{SciPy} and \texttt{matplotlib}. The software runs as a single application that keeps track of the GUI, the instrument connections, and the measurement data together. Its work can be grouped into four parts: instrument control, measurement sequencing, on-line analysis, and data handling. Measurement sequencing covers the single-channel and batch acquisition steps, along with the voltage-ramping and safety-interlock logic. On-line analysis covers the breakdown-voltage and quenching-resistance measurements, and data handling covers real-time plotting, auto-saving, and macro-based configuration reload.

The software communicates with a Keithley 2410 source-measure unit (SMU) over a VISA and SCPI interface, with automatic instrument discovery and a manual address selection dialog as a fallback when auto-detection fails. Voltage biasing is implemented as a software-controlled ramp with independently configurable step size and inter-step delay for ramp-up, ramp-down, and single-point bias changes. A user-defined current-compliance limit protects the device under test throughout the sweep, and the instrument output can be enabled, disabled, or reset to zero bias at any point during operation.}

\changed{The GUI software facilitates automated, sequential current-voltage (I-V) measurements of multiple SiPMs using dedicated readout electronics (detailed in a subsequent section). Furthermore, the data acquisition process is fully controllable, allowing users to pause, resume, or terminate operations at any point during both single-device and batch measurements. The GUI stays responsive during long voltage ramps and inter-channel delays because these operations are broken into short steps. A built-in ``Simulate'' button in the GUI allows users to run the application and visualize I–V data without connected hardware, enabling easy offline testing and demonstrations. A brief overview of the application workflow is shown in figure~\ref{fig:flowchart}. The full source code is publicly available at \cite{dey_pysipmgui_2026}.

\begin{figure}[H]
\centering\includegraphics[width=0.7\textwidth]{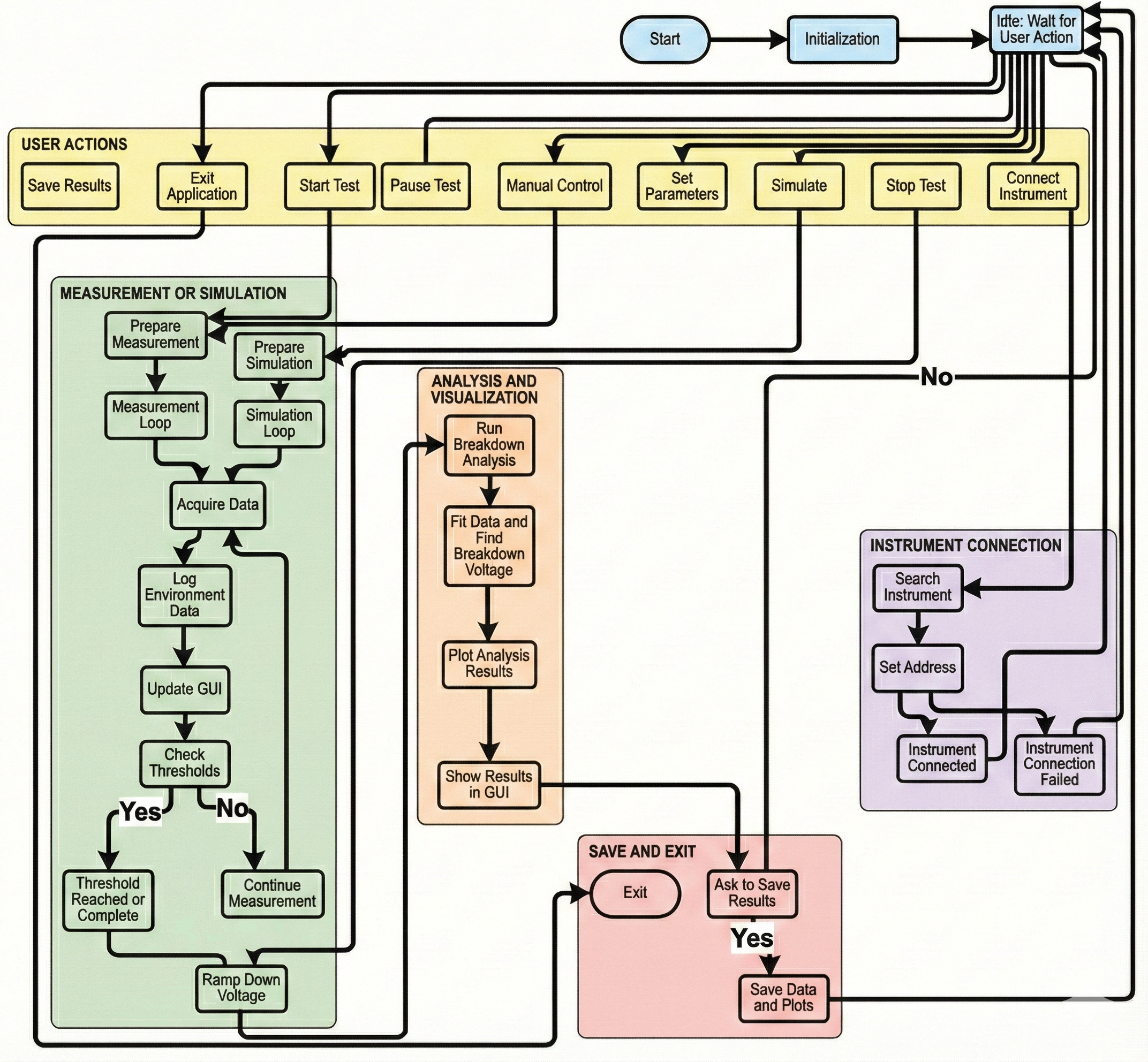}
\caption{\changed{Workflow of PySiPMGUI.} \label{fig:flowchart}}
\end{figure}

The remaining parts of the software listed above are described in the following subsections.}

\subsection{Safe Control Framework and Environmental Monitoring}
\label{sec:safety}

The SiPM characterization framework includes essential safety controls. Users can set voltage and current limits before applying bias to the SiPMs, which helps prevent accidental overstressing. These safety features are discussed further in the following sections.

\subsubsection{Voltage Control and Safe Ramping Algorithm}

The bias voltage is provided by a programmable source-measure unit (SMU) (e.g Keythley 2400 series). It can be operated in voltage-source mode with current compliance enabled. Sudden voltage changes can damage the SiPM. Therefore, a recursive voltage ramping routine, \texttt{ramp\_up\_run}, is used, which causes the voltage to increase in small steps.

For a target voltage $V_{\text{target}}$, the next intermediate setpoint is calculated as

\begin{equation}
    V_{\text{next}} = V_{\text{current}} \pm V_{\text{step}} .
\end{equation}
\changed{Here, $V_{\text{current}}$ denotes the voltage presently applied at the output of the SMU, as read back from the instrument.}
To prevent overshoot, the ramping algorithm checks the polarity of the voltage against the target voltage at every step and before a new voltage is applied. If $V_{\text{next}}$ goes beyond the target value, the output is immediately clamped to $V_{\text{target}}$. A stabilization delay time $\tau_{\text{wait}}$ is applied to settle the voltage at the output of the SMU and the device under test (DUT). If the target voltage is still not reached, the function schedules itself again after $\tau_{\text{wait}}$.

\subsubsection{Safe Ramp-Down Protocol}

When the current reaches its safety limit, the ramp-down process starts automatically to prevent  the damaged SiPMs from being subjected to large currents for an extended period. Therefore, the voltage is brought back to zero in a smooth and controlled manner. Some power supplies can deliver both negative and positive voltages. Near zero, polarity can be changed during ramping down, which can reverse the biasing mode of the SiPM. To prevent this, the step size is adjusted dynamically during the ramp-down and is finally set to zero. 
\begin{equation}
    V_{\text{step}}^{\text{dynamic}} =
    \begin{cases}
        V_{\text{step}}^{\text{user}} / 2, & |V_{\text{current}}| \le V_{\text{step}}^{\text{user}}, \\
        V_{\text{step}}^{\text{user}},     & \text{otherwise}.
    \end{cases}
\end{equation}
This strategy avoids inadvertent reverse biasing of the SiPM caused by coarse voltage steps close to zero.

\paragraph{Safety Interlocks and Compliance Handling.} \label{subsec:saftey}

Two independent safety interlocks are evaluated during every measurement cycle:
\begin{itemize}
    \item \textbf{Hardware Compliance Detection:} The requested source voltage ($V_{\text{source}}$) is compared with the measured output voltage ($V_{\text{meas}}$). A deviation satisfying
    \begin{equation}
        |V_{\text{source}} - V_{\text{meas}}| \ge V_{\text{step}},
    \end{equation}
    indicates that the SMU has entered over-current protection mode, effectively switching from voltage sourcing to current limiting.

    \item \textbf{Software Current Threshold Monitoring:} The measured leakage current ($I_{\text{meas}}$) is continuously compared against a user-defined safety limit ($I_{\text{th}}$). A preemptive stop is triggered when
    \begin{equation}
        |I_{\text{th}} - I_{\text{meas}}| \le 20\,\text{nA},
    \end{equation}
    providing a protective margin before the allowable current is exceeded.
\end{itemize}

If either condition is met, the system shows a warning message and automatically initiates \changed{the ramp-down sequence to safely reduce  the applied voltage to zero, thereby preventing damage to the SiPM sensor.} A brief overview of the flow of the code is shown as a flowchart in the figure \ref{fig:flowchart}.

\subsection{Temperature and Humidity Monitoring}

It is well established that the breakdown voltage of an SiPM is linearly dependent on the temperature. For long-term operation, variations in the ambient temperature can shift the breakdown voltage; hence, the set overvoltage also changes. As a result, it can reduce or increase the expected gain and photon detection efficiency of the SiPM. To mitigate this effect, an Arduino-based \cite{arduino_uno_r3_datasheet} temperature monitoring unit is added to the setup. When the ``Connect Arduino'' option is enabled on the GUI, communication with the Arduino is established using the \texttt{pySerial} library. \changed{Before each voltage step, a query is sent to the Arduino requesting all sensor readings.} The sensor readings are then read back, processed, and stored along with the I-V data. These temperature readings are later used to account for temperature-driven changes in the breakdown voltage ($\changed{V_\text{BD}}$). \changed{Any environmental sensor communicating with an Arduino via the I$^2$C protocol can be integrated by implementing the appropriate firmware. Provided the Arduino returns tab-separated temperature and humidity values upon request, the Python-side data parsing is handled generically. An example of the Arduino firmware can be found in the GitHuB repository \cite{dey_pysipmgui_2026}.}

\changed{\subsection{Acquisition modes}
\label{subsec:acquisition}

Two measurement modes are supported for each channel: forward-bias sweeps for quenching-resistance determination, and reverse-bias sweeps for extracting the breakdown voltage and dark count rate (DCR). Sweep parameters -- start and end voltage, step size, settling delay, and current-compliance limit -- are independently configurable per mode.

\subsection{On-line analysis}
\label{subsec:analysis}

Acquired I--V data are analyzed automatically as each sweep completes. The breakdown voltage $V_{BD}$ is extracted by fitting the reverse-bias branch to a Geiger-mode avalanche-gain model; fit quality is reported together with the associated parameter uncertainties. The quenching resistance $R_q$ is extracted by applying a linear fit to the ohmic region of the forward-bias I--V curve. Results are displayed on live, auto-scaling plots (with automatic selection of appropriate current and voltage units) supporting linear or logarithmic axis scaling, error-bar display, and simultaneous dual-axis presentation of auxiliary environmental data (e.g.\ temperature and humidity) alongside the primary I--V trace.

\subsection{Fault detection and recovery}
\label{subsec:fault-recovery}

The software monitors instrument communication for signatures of a lost USB connection during an active sweep. On detection of a disconnection, the affected measurement is safely suspended, and the application attempts automatic reconnection to the instrument; upon successful reconnection, the sweep is resumed from its last completed point rather than restarted, minimizing data loss during unattended or batch operation.

\subsection{Data management and post-processing}
\label{subsec:data-management}

Raw voltage/current data, derived fit parameters, and generated plots are autosaved automatically for each channel during batch operation, and can additionally be saved interactively during single-channel measurements. Complete measurement and batch configurations -- sweep parameters, channel assignments, and analysis settings -- can be exported to and reloaded from macro files, enabling exact reproduction of an acquisition sequence without manual GUI reconfiguration.}

\section{ Theoretical Modeling of SiPM I-V Characteristics}
\label{sec:theory}

It has already been discussed that to operate a SiPM at a fixed gain, we need to operate it at a fixed overvoltage, which can be achieved when we correctly determine the breakdown voltage $V_{BD}$. The breakdown voltage $V_{BD}$ can be obtained by fitting the reverse I-V data with the physical model suggested by ~\cite{DINU_1}, which we adopt in preference to the logarithmic-derivative method, as the latter lacks a strong physical basis and can be disturbed by parasitic or leakage currents. The proposed dark current model ~\cite{DINU_1} explains the full I-V curve over the entire voltage range at a fixed temperature by directly including Geiger discharge and afterpulsing in the SiPM. The total current can be modeled as the sum of the pre-breakdown leakage current and the post-breakdown avalanche current.
\subsection{Pre-Breakdown Leakage Current ($I_{leak}$)}
According to the model ~\cite{DINU_1}, in the region where $V < V_{BD}$, the current is dominated by surface leakage and bulk thermal carriers. This is parameterized empirically as an exponential function, 
\begin{equation} \label{eq:leak}
    I_{leak}(V) = \exp(a \cdot V + b),
\end{equation}
where $a$ and $b$ are fit parameters describing the baseline dark current independent of the avalanche gain.

\subsection{Post-Breakdown Avalanche Current ($I_{aval}$)}
In the Geiger mode ($V > V_{BD}$), the current is driven by free carriers triggering avalanches. The model describes this current as a product of the primary carrier rate, the charge per pulse, the triggering probability, and an afterpulsing amplification factor. The current is given by ~\cite{DINU_1}:
\begin{equation}
    I_{aval}(V) = \frac{dN_{carriers}}{dt} \cdot C_{\mu cell} \cdot (V - V_{BD}) \cdot P_{Geiger}(V) \cdot G_{AP}(V)
\end{equation}
where:
\begin{itemize}
    \item $\frac{dN_{carriers}}{dt}$ is the rate of thermally generated free carriers (primary dark count rate).
    \item $C_{\mu cell}$ is the microcell capacitance.
    \item $P_{Geiger}(V) = 1 - \exp(-p \cdot (V - V_{BD}))$ is the Geiger triggering probability, with $p$ representing the probability shape factor.
    \item $G_{AP}(V) = \frac{V_{cr} - V_{BD}}{V_{cr} - V}$ is the gain factor due to afterpulsing, which diverges as the voltage approaches a critical value $V_{cr}$ where afterpulsing becomes self-sustaining.
\end{itemize}

Combining these terms leads to the full fit equation used in this work:
\begin{equation}\label{eq:idark}
    I_{tot}(V) = \exp(aV + b) + A \cdot (V - V_{BD}) \cdot \left(1 - e^{-p(V - V_{BD})}\right) \cdot \frac{V_{cr} - V_{BD}}{V_{cr} - V}
\end{equation}
Here, the amplitude parameter $A$ groups the constant terms ($A \approx \frac{dN_{carriers}}{dt} \cdot C_{\mu cell}$).
\subsection{Temperature Dependent SiPM Dark Current} \label{sec:temp_dark}
It is known that the Geiger probability becomes non-zero when the bias voltage is greater than the breakdown voltage $V_\text{BD}$. Therefore, the total dark current of SiPM in the \changed{Geiger region (i.e., the bias voltage regime above the breakdown voltage, $V > V_\text{BD}$, where each thermally generated carrier has a non-zero probability of triggering a self-sustaining avalanche discharge)} mainly arises from the thermally generated charge carriers that trigger Geiger Avalanches. Since the charge carriers generated inside the active depletion region are amplified by \changed{the large avalanche multiplication factor (SiPM gain $G \sim 10^6$)}, the other components of total current, such as surface leakage and peripheral current, become ineffective. The measured current is proportional to the total charge in the avalanche and the rate of triggering events, which can be written as below:
\begin{equation}
I_{\mathrm{dark}}(T) = q \, G(T) \, \mathrm{DCR}(T),
\end{equation}
where, $q$ is the elementary charge, $G(T)$ is the SiPM gain, and $\mathrm{DCR}(T)$ is the dark count rate. 

The bulk dark current arises from two thermally activated processes. The first process is Shockley-Read-Hall (SRH) generation in the depleted region \cite{srh1,srh2,Sze_Ng_Physics_Semiconductor_Devices}. SRH generation is mediated by deep defect states in the silicon bandgap and dominates at low temperatures, where the intrinsic carrier concentration is small. The second process is the diffusion of thermally generated carriers from the quasi-neutral regions into the high-field multiplication region \cite{Sze_Ng_Physics_Semiconductor_Devices}. In both carrier-generation processes, the dark current can be modeled as \cite{Biroth_I_dark_vs_T},

\begin{equation} \label{eq:arr_eq}
I_{\mathrm{SRH/Diff}}(T) = A_{\mathrm{SRH/Diff}}^\prime \, T^{3/n}
\exp\!\left(-\frac{E_{\mathrm{eff}}}{n\,k_B T}\right),
\end{equation}
where, $A_{\mathrm{SRH/Diff}}^\prime$ is a temperature-independent prefactor and $E_{\mathrm{eff}}$ is the the effective band-gap of the silicon, which is approximately 1.12 eV at room temperature. The value of $n \approx 2,1$  in SRH and diffusion region, respectively.

It is known that at a fixed overvoltage, the SiPM gain decreases linearly with temperature. Therefore, it can be written as \cite{Eigen_2019,Jangra_2022},

\begin{equation}
G(T) = \alpha \left(1 - \beta (T-T_0) \right),
\end{equation}

where, $\alpha$ is a proportionality constant and $\beta$ is the fractional gain temperature coefficient and $T_0$ is the reference temperature $\approx$ 300 K. Combining the temperature dependence of the gain with the bulk generation mechanisms, the total dark current can therefore be written as,

\begin{equation}\label{eqn:I_dark_full_1}
I_{\mathrm{dark}}(T) = q \left(1 - \beta (T-T_0) \right)
\left[
A_{\mathrm{SRH}}^\prime \, T^{3/m} \exp\!\left(-\frac{E_{\mathrm{eff}}}{m\,k_B T}\right)
+
A_{\mathrm{Diff}}^\prime \, T^{3/n} \exp\!\left(-\frac{E_{\mathrm{eff}}}{n\,k_B T}\right)
\right],
\end{equation}
where, the constant term  and $\alpha$ are absorbed in the \changed{$A'_{\mathrm{SRH/Diff}}$}.

As $A_{\mathrm{SRH/Diff}}^\prime$ gives a large number hence to get numerical stability the equation \ref{eqn:I_dark_full_1} can be wirtten as :
\begin{equation}\label{eqn:I_dark_full}
I_{\mathrm{dark}}(T) = q \left(1 - \beta (T-T_0) \right)
\left[
\, T^{3/m} \exp\!\left(A_{\mathrm{SRH}}-\frac{E_{\mathrm{eff}}}{m\,k_B T}\right)
+
\, T^{3/n} \exp\!\left(A_{\mathrm{Diff}}-\frac{E_{\mathrm{eff}}}{n\,k_B T}\right)
\right],
\end{equation}
where, $A_{\mathrm{SRH/Diff}}^\prime=\exp(A_\mathrm{SRH/Diff})$.

\begin{figure}[h]
    \centering\subfloat[\label{fig:box_photo}]{
        \includegraphics[width=0.45\linewidth]{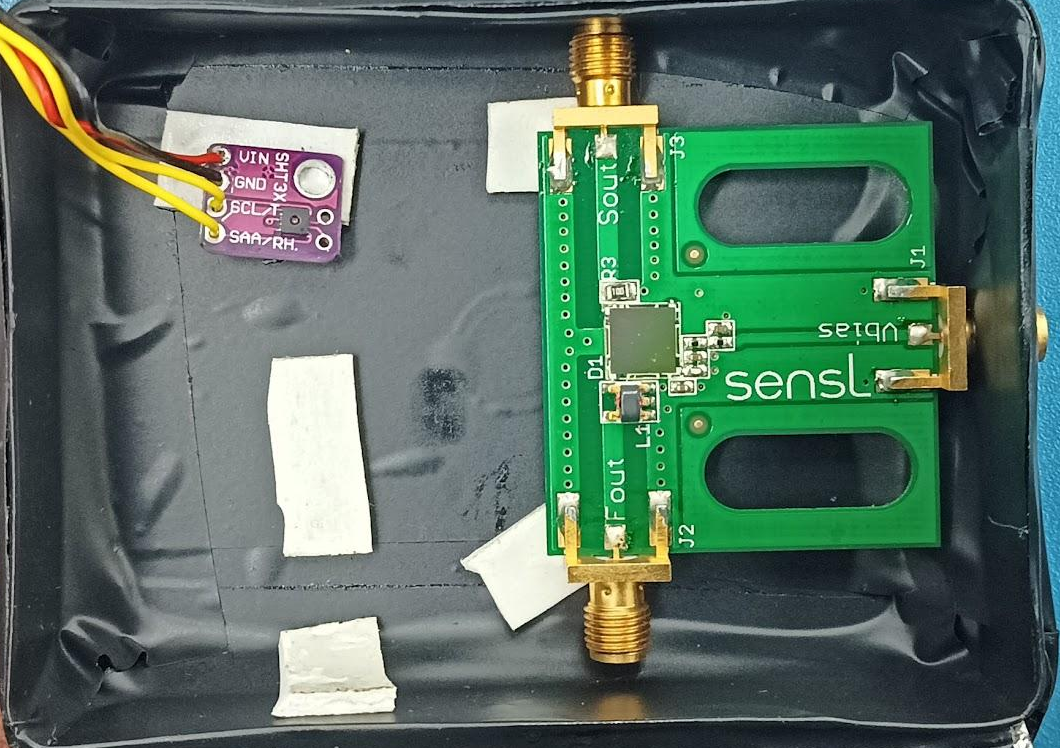}
        \label{fig:sipm_photo}
    }
    
    \centering\subfloat[\label{fig:bias_circuit}]{
        \includegraphics[width=0.95\linewidth]{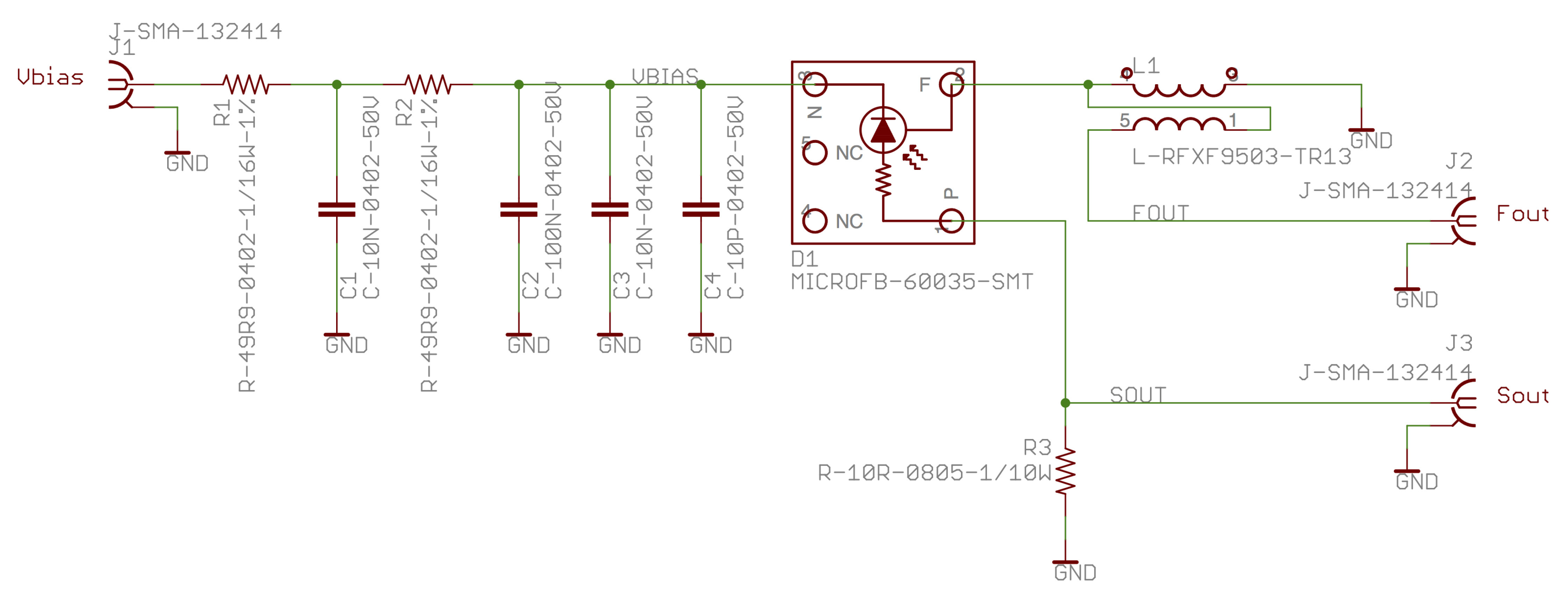}
        \label{fig:sipm_bias}
    }
    \caption{(a) SiPM device and experimental setup used for characterization; (b) Biasing Circuit Design for MicroFC-60035-SMT\cite{onsemi_AND9809_2018}.}
    \label{fig:sipm_setup}
\end{figure}
 \section{Experimental Methodology}
\label{sec:setup}
 \begin{figure}[h]
    \centering

    \includegraphics[width=0.95\textwidth]{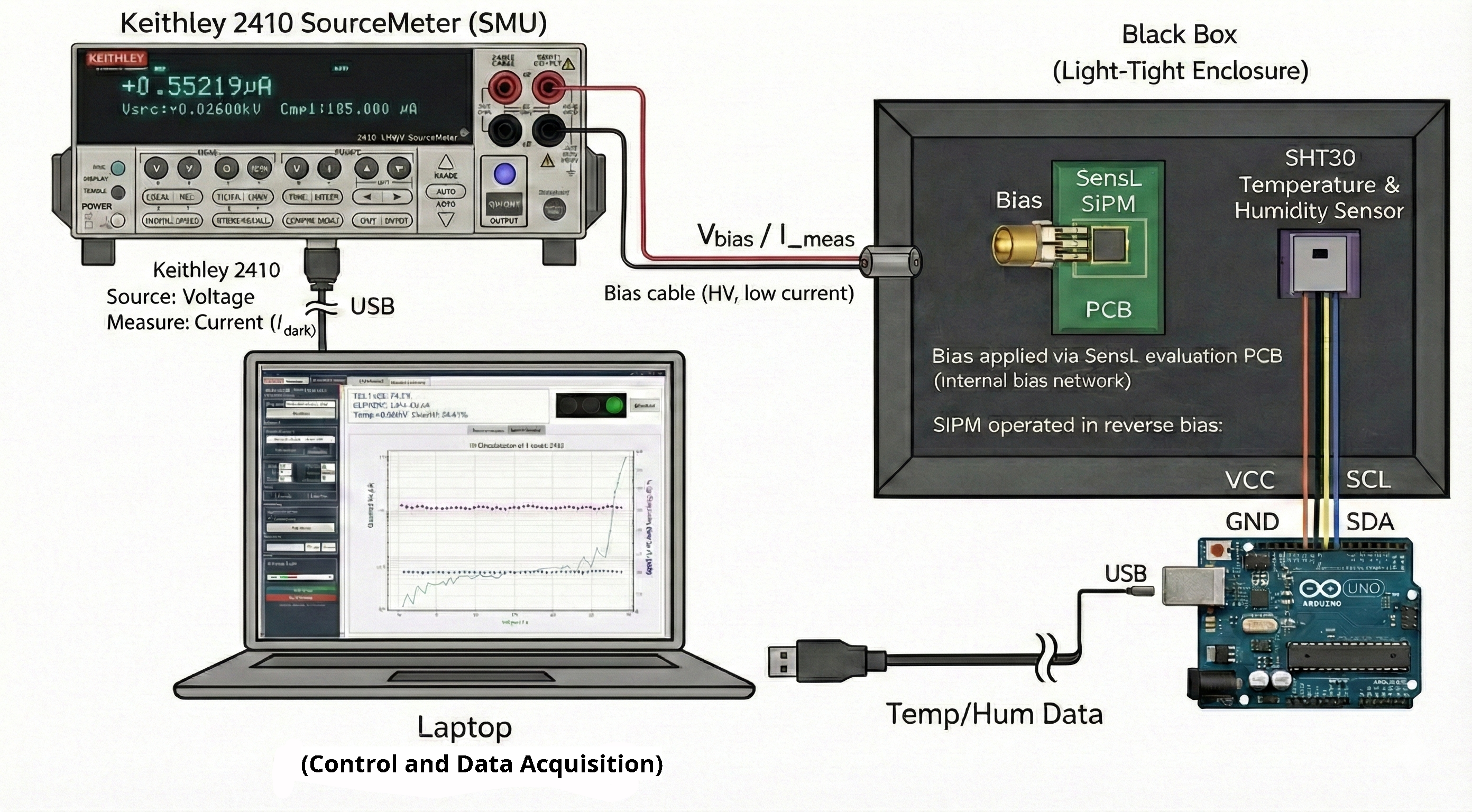}
    \caption{Schematic diagram of experimental setup} \label{fig:schematic}
    
\end{figure}

\subsection{\changed{Sensor Specifications}}\label{sec:detector_description}
The SiPM used in this study is the onsemi (SensL C-Series) MicroFC-60035-SMT~\cite{onsemi_sipm_datasheet_2022}, a ready-made module \changed{(MICROFC-SMA-60035-GEVB)} with an integrated bias circuit, as shown on the green PCB board in figure~\ref{fig:box_photo}. According to the datasheet \cite{onsemi_sipm_datasheet_2022}, the same SiPM has an active area of $6 \times 6$ $mm^2$ and consists of 18,980 microcells with a microcell size of 35 $\mu m$ and a fill factor of 64\%. It is sensitive to UV-visible light in the wavelength range of 300-900 nm and exhibits a peak photon detection efficiency at 420 nm. At an overvoltage of 2.5 V, the SiPM has a gain of approximately $3 \times 10^6$; the breakdown voltage at 21 $^\circ\mathrm{C}$ is ranged 24.2-24.7 V, and the dark count rate is about 1.2-3.4 MHz. 
\subsection{Biasing and Readout Circuitry}
The biasing circuit used for the measurements is shown in figure \ref{fig:bias_circuit}, which is manufacturer-made \cite{onsemi_AND9809_2018}. A passive low-pass filtering stage was implemented on the bias voltage line ($V_{\text{bias}}$) to reduce high-frequency noise originating from the power supply. This filter comprises a multi-stage RC network, with series resistors $R_1$ and $R_2$ of $49.9~\Omega$ and shunt capacitors $C_1$ and $C_3$ of $10$~nF, along with an additional $100$~nF capacitor $C_2$, placed upstream of the SiPM cathode.

The SiPM anode is connected to the standard output ($S_{\text{out}}$) through a $10~\Omega$ load resistor ($R_3$), which provides the DC current path used for current--voltage measurements. In addition, a fast output ($F_{\text{out}}$) is available through capacitive coupling via an RF transformer ($L_1$) for timing-related studies. 
\subsection{Experimental Setup}
 An experimental setup has been designed to measure the dark current of SiPMs by varying the bias voltage. \changed{As described in the preceding sections, the bias voltage sweep and current measurements were fully automated using PySiPMGUI, employing a Keithley~2410 SourceMeter as the SMU.} 
A schematic overview of the complete measurement system is shown in figure \ref{fig:schematic}.

To protect the SiPM from ambient light, it is kept inside a light-tight box as shown in the figure \ref{fig:box_photo}. In the same light-tight box (see figure \ref{fig:box_photo}) an SHT30 \cite{sensirion_sht3x}  digital sensor was installed in close proximity to the SiPM to log the ambient temperature and relative humidity. The SHT30 sensor interfaces via the I$^2$C protocol with an Arduino micro-controller \cite{arduino_uno_r3_datasheet}, which transmits the temperature and humidity data to the host PC via USB, as shown in the figure \ref{fig:schematic}. \changed{The complete Arduino firmware for sensor integration is available in the GitHuB repository \cite{dey_pysipmgui_2026}.
}
It should be noted that users may employ different temperature and humidity sensors. 

 \begin{figure}[h]
    \centering
        \subfloat[\label{fig:bias_distribution}]{\includegraphics[width=0.5\textwidth]{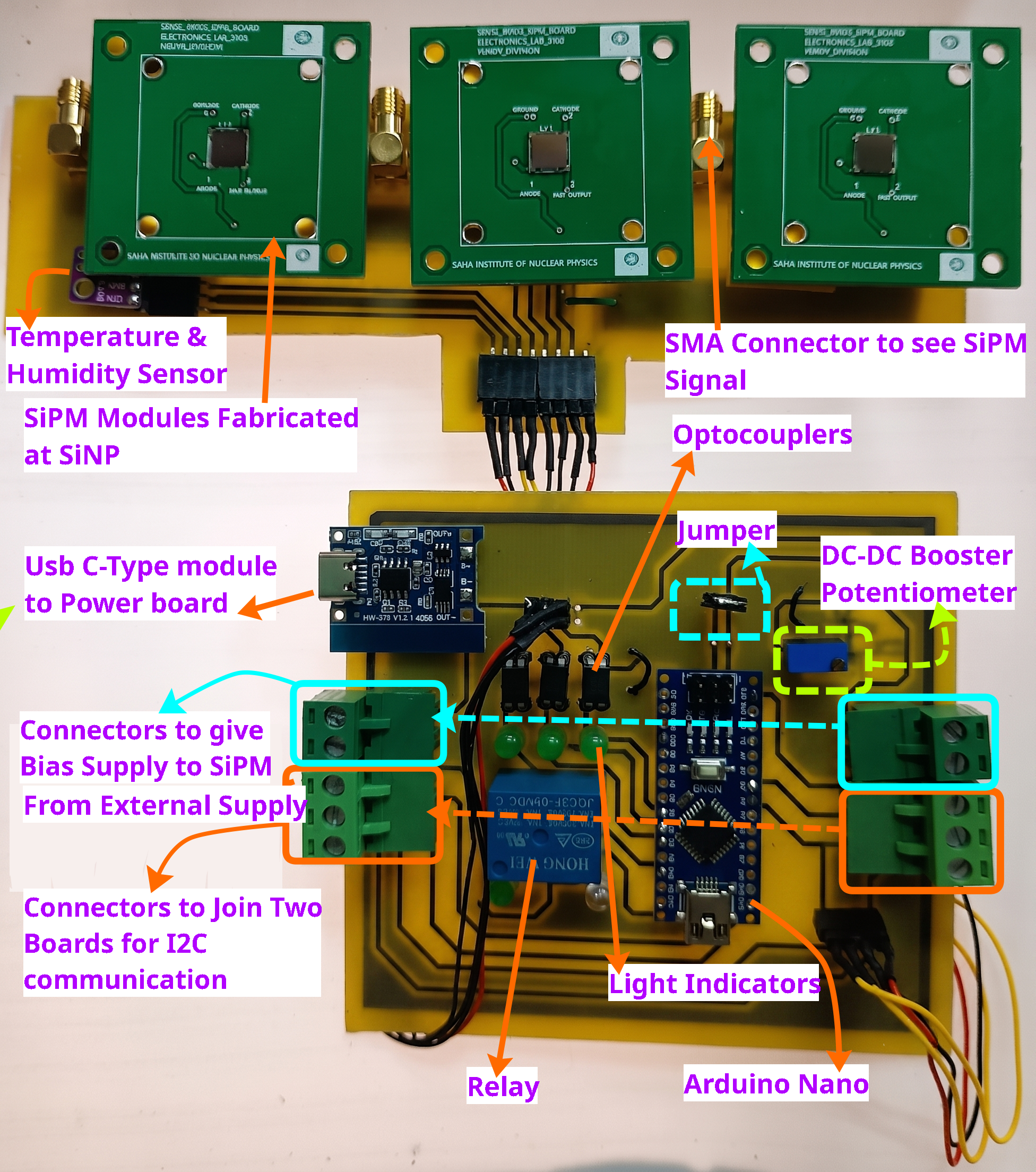}}
    
    \subfloat[\label{fig:bias_distribution_block}]{\includegraphics[width=0.5\textwidth]{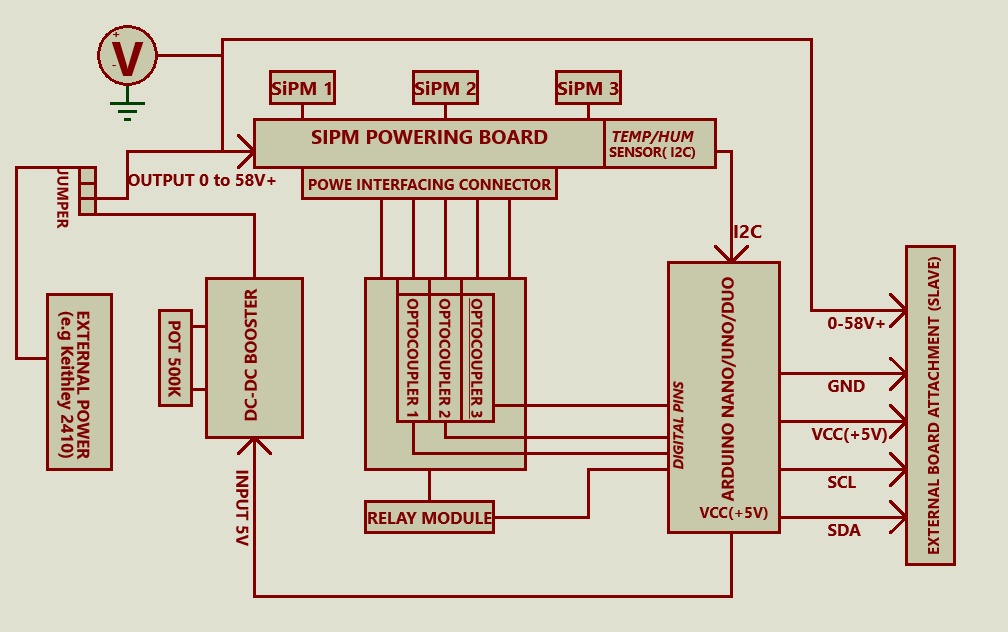}}

    \caption{Multi-channel bias voltage distribution board, (a) The original bias distribution circuit, (b) Block diagram of the power distribution board. }
\end{figure}

\subsection{Multi-Channel SiPM Bias Distribution Board}
\label{sec:bias_distribution}
 
\changed{To extend the single-channel readout described above to multiple SiPMs, two prototypes of a dedicated multi-channel bias distribution board were developed. The original circuit board and its block diagram are shown in figures~\ref{fig:bias_distribution} and \ref{fig:bias_distribution_block}, respectively. Currently, during I-V characterization, an external power supply (e.g., a Keithley 2410) is used to distribute a common high-voltage bias to the SiPMs. The board allows each channel to be switched on and off independently via software control, combining relay-based and optocoupler-based switching to ensure electrical isolation and safe operation during measurements. Alternatively, each board features an integrated DC-DC booster that can directly supply the bias voltage to its connected SiPMs. The required voltage for this booster can be adjusted using a potentiometer, and a jumper is provided to select between the booster and an external power supply, as shown in figures~\ref{fig:bias_distribution} and \ref{fig:bias_distribution_block}.}
 
\changed{Each board carries one relay, which gates the main SiPM bias input to the board, and a bank of optocouplers (three in the current design, one per SiPM slot), each of which independently enables or disables the bias delivered to a single SiPM channel. The relay and optocouplers are driven directly from Arduino digital output pins, with switching commands issued from the GUI; an on/off indicator LED is provided alongside each optocoupler for a quick visual check of channel state. The number of optocouplers -- and hence the number of SiPM channels -- per board is limited only by the number of available Arduino digital pins, and can be increased accordingly. 
A dedicated extender port and communication pins facilitate the daisy-chaining of multiple boards via an I$^2$C bus. This architecture allows a single master Arduino controller to address more slave boards than its available direct digital outputs would normally permit, effectively expanding the total channel capacity. The current setup implements two such boards, each featuring three SiPM slots for a combined six channels, managed by a single Arduino. Furthermore, this I$^2$C-based extension scheme provides inherent scalability, enabling the seamless integration of additional boards as channel requirements increase.
}
 
\changed{Two modes of operation are supported. During automated I--V characterization, channels are selected sequentially: the relay on the addressed board is switched on first, after which its optocouplers are enabled one at a time, so that only a single SiPM is biased at any given moment. Alternatively, when sequential characterization is not required, all channels across the connected boards can be enabled simultaneously, providing continuous bias to the full array, for example, during long-term monitoring or detector operation rather than dedicated I--V scans.}
 
\changed{This design is at the prototype stage. Two boards have been fabricated and tested, each populated with three SiPMs, for a total of six channels biased and read out through the system described above. The modular, I$^2$C-addressable architecture is intended to scale to larger detector arrays and higher-throughput automated testing without changes to the underlying control software.}

\section{\changed{Automated I-V Characterization Logic}}
\label{sec:iv_logic}

\changed{The I-V characterization is performed in several steps via the GUI, which enables automated control of both single-channel and multi-channel batch measurements using PySiPMGUI. The two modes of operation are described in turn below.}

\subsection{Single-Channel I-V}
\begin{figure}[htbp]
    \centering
    \includegraphics[width=1.\textwidth]{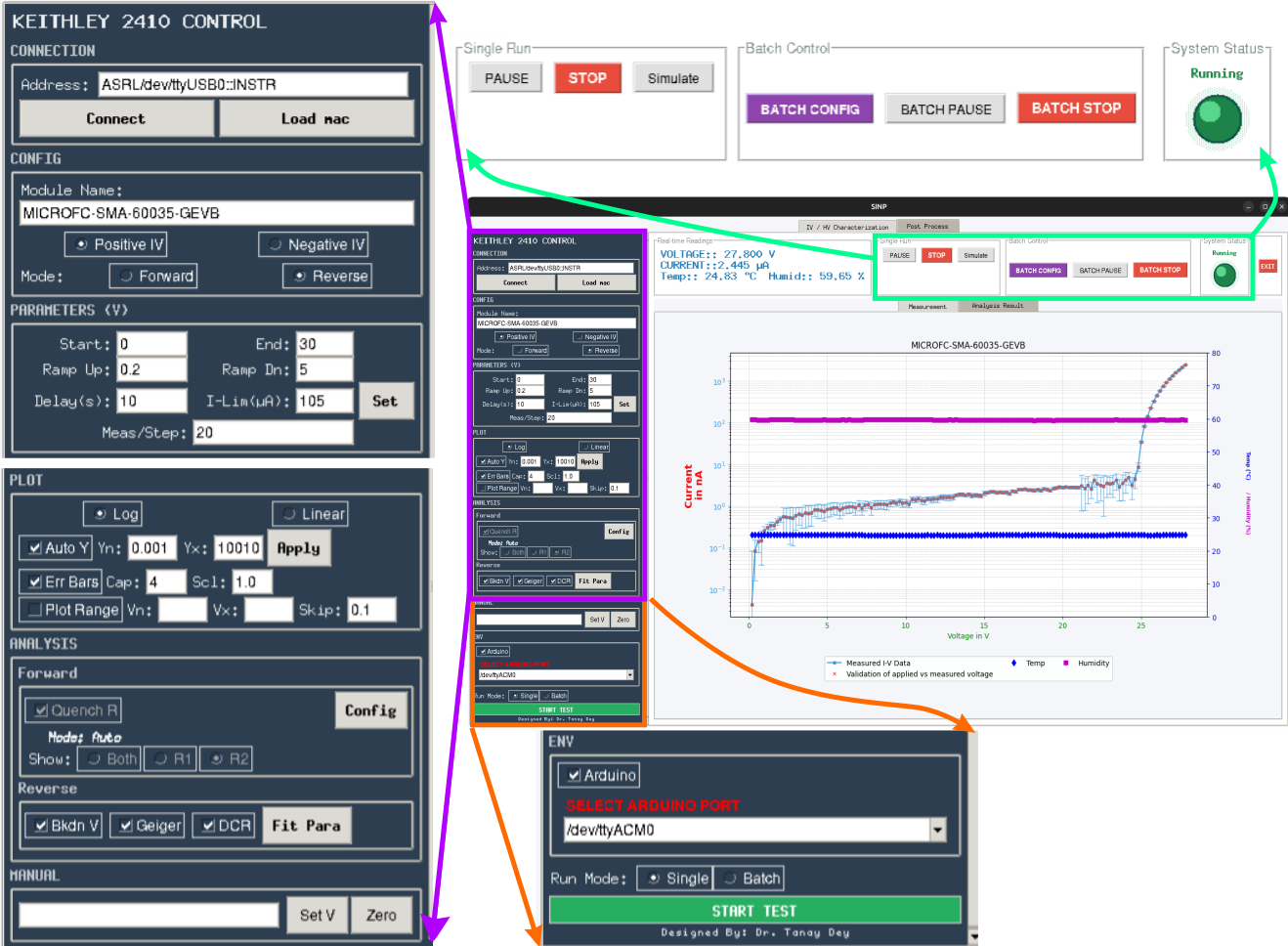}
    \caption{\changed{The Python-based GUI during an I-V characterization run with humidity and temperature monitoring. The middle line represents the average current (in nA) at any given voltage, the pink line represents humidity, and the blue line represents temperature.  The left sidebar contains configuration controls for voltage range, step size ($V_{step}$), and compliance limits.}}
    \label{fig:gui_screenshot}
\end{figure}

To measure I-V, the user first needs to select the polarity of the output voltage (positive or negative), and then the I-V mode (forward or reverse), as shown in the left panel of figure \ref{fig:gui_screenshot}. In the next step, the voltage sweep parameters,  such as start and end voltage, step size, and delay between measurements, as well as the maximum current limit to be defined on the GUI (see figure \ref{fig:gui_screenshot}). The desired plot scale (linear or logarithmic) is also chosen at this stage. \changed{The user can specify the number of measurements, $N$, taken at each voltage step. The resulting plot displays the mean current, using the standard deviation to indicate uncertainty via error bars}. To show and log temperature and humidity data on the plot, the ``Connect Arduino'' option can be checked.  Once the connection is confirmed, the test is started using the ``START TEST'' button. An example of measured reverse I-V data are displayed in real time on the selected scale in the middle panel of the figure~\ref{fig:gui_screenshot}, allowing the user to identify key features such as breakdown behavior and leakage current.

\begin{figure}[htbp]
    \centering
    \includegraphics[width=0.85\textwidth]{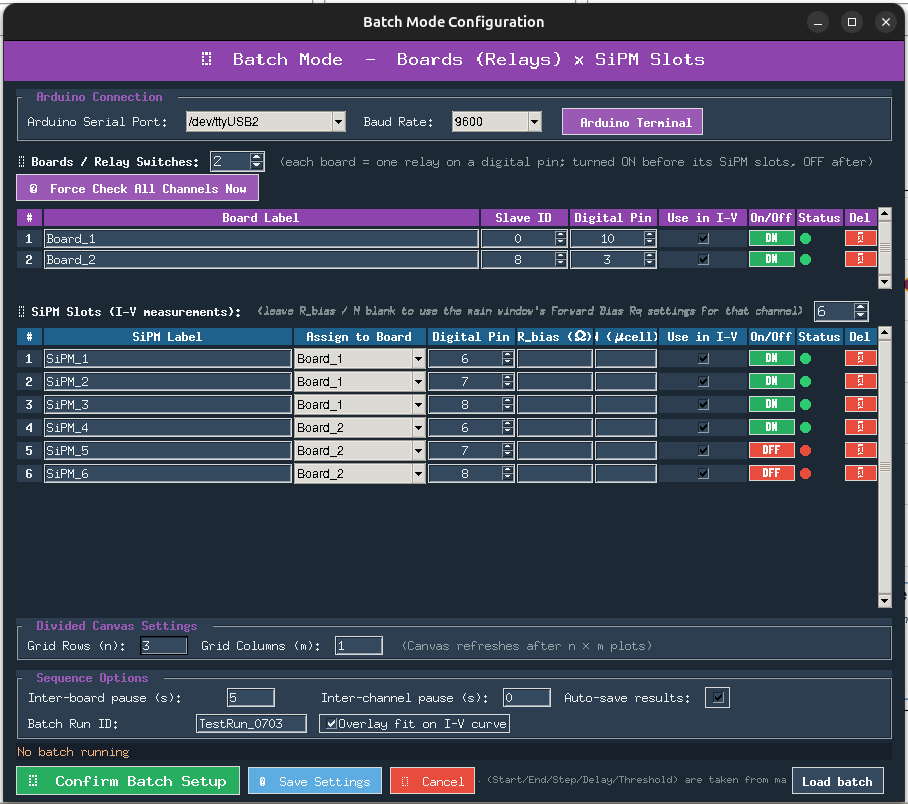}
    \caption{\changed{Batch Mode Configuration panel of PySiPMGUI. The Arduino connection settings, the list of configured relay boards, and the list of SiPM slots (with per-channel assignment to a board, digital pin, and bias resistance/microcell capacitance) are shown, along with sequence options such as inter-board and inter-channel pause times and the batch run ID.}}
    \label{fig:batch_config}
\end{figure}
\changed{\subsection{Multi-Channel I-V}

Selection of individual SiPM channels for batch operation is performed through an Arduino-based voltage distribution board as discussed in section \ref{sec:bias_distribution}. The software lets users configure a custom number of relay boards and channels. As shown in figure \ref{fig:batch_config}, it automatically creates controls to map channel IDs to specific board pins. Each channel can be individually enabled, disabled, or queried for its current relay state, and per-channel status indicators provide real-time visual feedback (idle, running, warning, error, or completed) during automated operation. To extend the single-channel procedure, a batch sequencer automates the entire multi-channel acquisition process. It steps through all configured boards and channels in sequence, automatically handling hardware switching, delays, and sweep execution based on global or overridden parameters. To extend the single-channel procedure, a batch sequencer automates the entire multi-channel acquisition process. It steps through all configured boards and channels in sequence, automatically handling hardware switching, delays, and sweep execution based on global or overridden parameters. Pause, resume, and immediate-stop control of batch acquisition is available at any stage of the sequence.
\begin{figure}[htbp]
    \centering
    \includegraphics[width=0.95\textwidth]{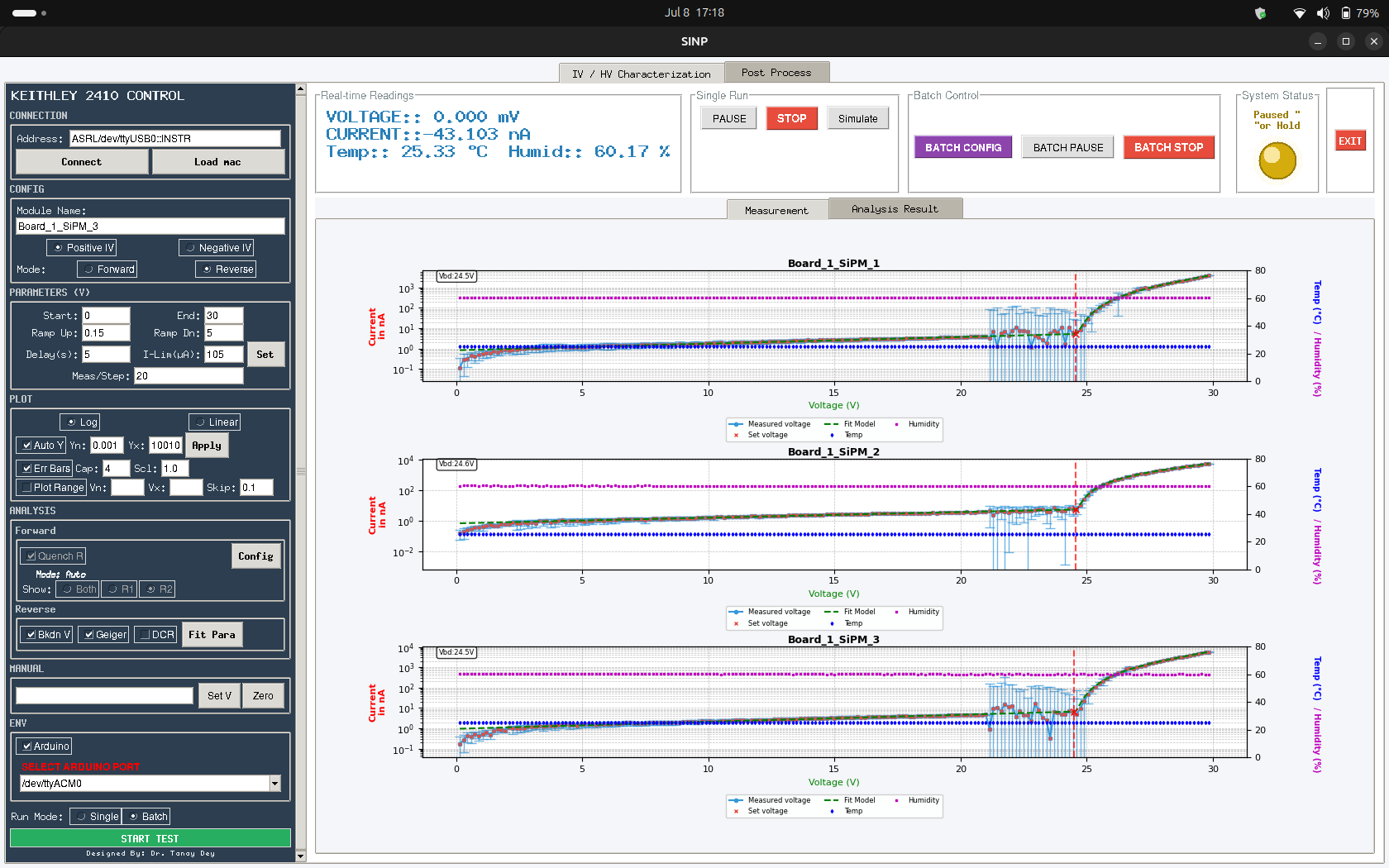}
    \caption{\changed{The Python-based GUI during an I-V characterization run. The left sidebar contains configuration controls for voltage range, step size ($V_{step}$), and compliance limits. The central monitoring strip displays real-time scalar values for voltage, leakage current, and environmental data (temperature/humidity).}}
    \label{fig:multi_sipm_i-V_GUI}
\end{figure}
}
\changed{A batch run can involve many channels. Because of this, users can customize the layout of the live I--V plots. They simply enter the desired number of rows and columns in the configuration panel. The software then divides the canvas into that exact number of sub-plots. As each channel is measured, its I--V curve is drawn in the next free sub-plot of the grid. Once all the sub-plots on the canvas are filled, the canvas is cleared and redrawn with a fresh grid for the next set of channels, so the display cycles through the batch in pages rather than trying to show every channel at once. This lets the user choose a layout of $m\times n$ grid, that best fits the number of channels being tested and the size of the screen. An example of three SiPMs measured sequentially in batch I--V mode is shown in figure~\ref{fig:multi_sipm_i-V_GUI}. After each I--V scan, the breakdown-voltage analysis is performed automatically and the results are saved to a separate folder.}

\section{Results and Validation}
\label{sec:results}
\subsection{Determination of Breakdown Voltage ($V_{BD}$)}
\changed{\subsubsection{Results of Readymade Module MICROFC-SMA-60035-GEVB}}\label{sec:results_readymade}

 \begin{figure}[h!]
    \centering
    \subfloat[\label{fig:break_option}]
    {\includegraphics[width=0.8\textwidth]{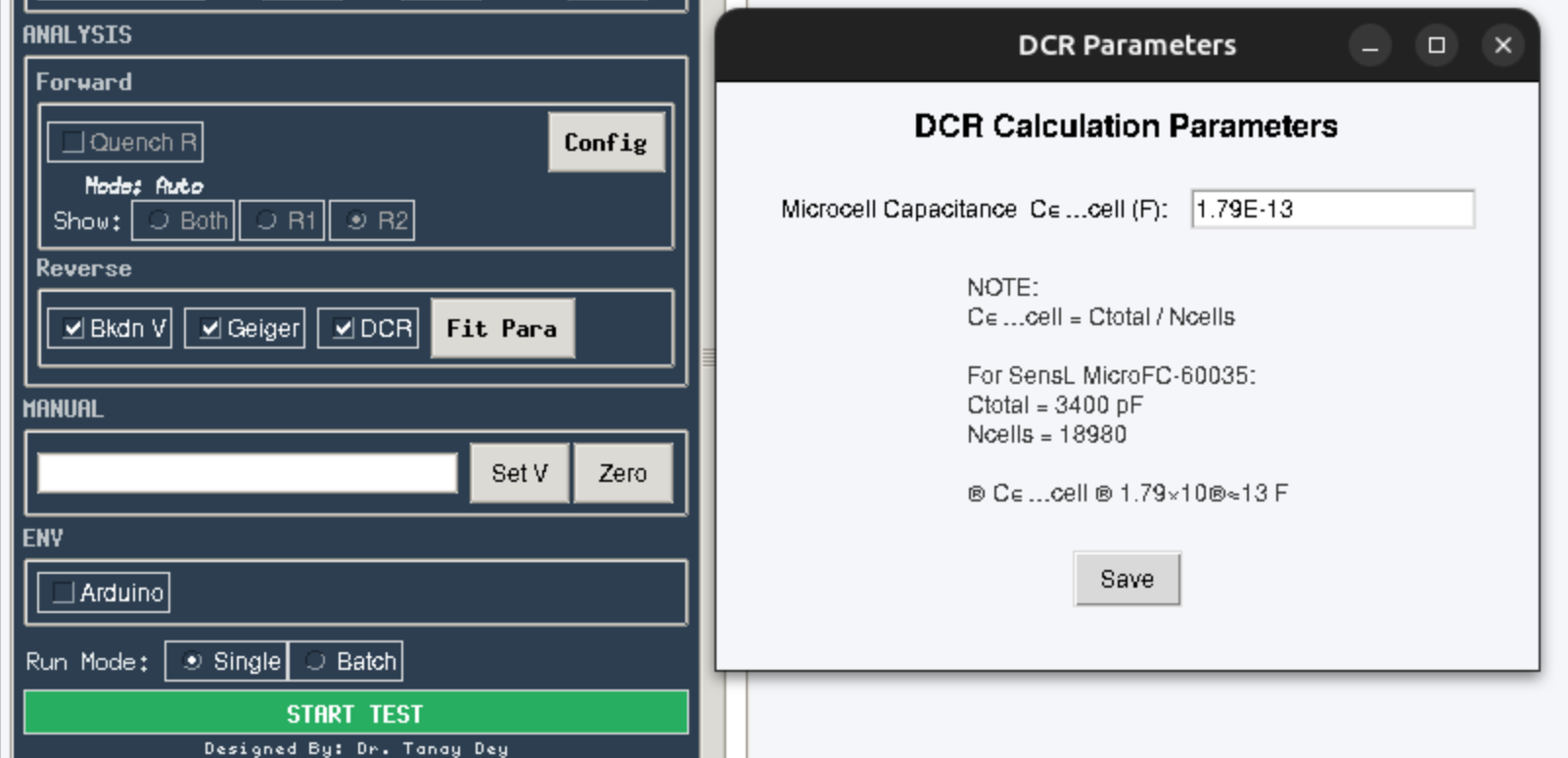}}

    \subfloat[\label{fig:post_option}]
    {\includegraphics[width=0.9\textwidth]{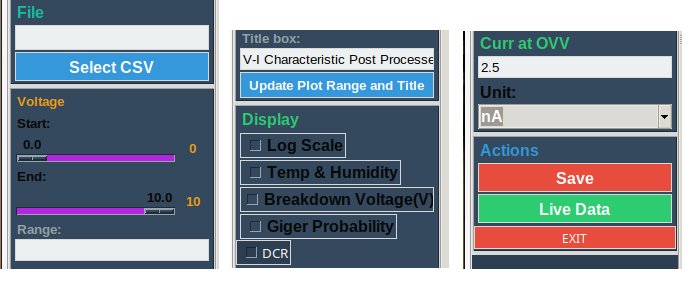}}
     \caption{(a) Pannel to enable breakdown voltage and dark count rate (DCR) analysis. (b) Snapshot of pannel for post-processing. \label{fig:dcr_gui_screenshot}}   
\end{figure}
 \begin{figure}[htbp]
    \centering
    \subfloat[\label{fig:break_graph_fit}]{\includegraphics[width=0.75\textwidth]{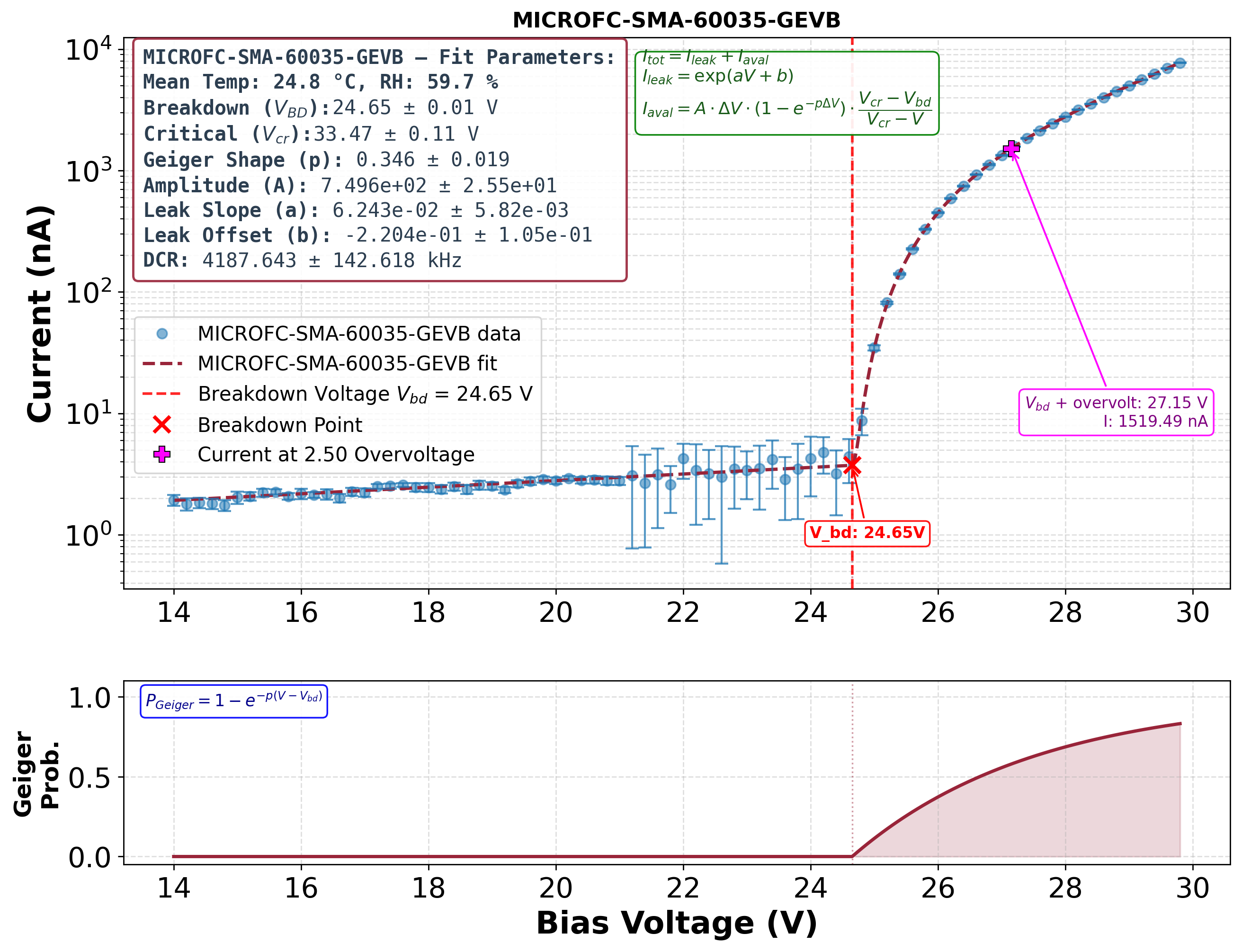}}
    
    \subfloat[\label{fig:data_sheet}]{\includegraphics[width=0.6\textwidth]{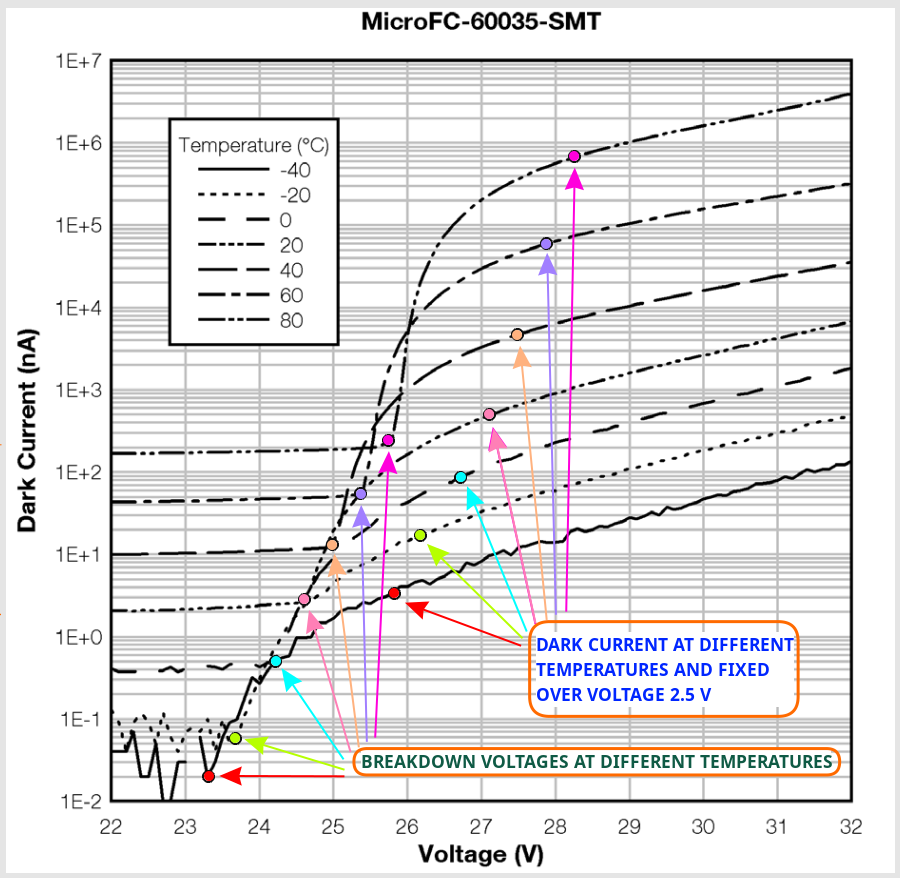}}
    \caption{(a) Determination of the breakdown voltage and other parameters by fitting the I--V data with equation ~\ref{eq:idark}; 
(b) The values of the breakdown voltage and the current at an overvoltage of 2.5~V are taken from the I--V data at different temperatures from the manufacturer's datasheet.
\cite{onsemi_sipm_datasheet_2022}\label{fig:gui_screenshot_fit}}
    
\end{figure}
To enable breakdown analysis, the user needs to check the option Breakdown, as shown in figure \ref{fig:break_option}. After finishing the data collection, the $I-V$ data are fitted to the equation \ref{eq:idark} using the least-square method available in Python as shown in figure \ref{fig:break_graph_fit}, and the fitted graph is shown in a tab on the GUI named ``Analysis result''. There is also a ``Post Process'' tab, where the user can further analyze the data using a suitable scale and range as shown in figure \ref{fig:post_option}. To fit the data, the user can set initial parameters using the button ``Fit Para''. If no parameter is provided, the code automatically chooses the initial $V_\text{BD}$ by calculating $\frac{d\log I}{dV}$. Then the initial $V_\text{BD}$ can be determined from the location of the maximum of  $\frac{d\log I}{dV}$, let's say $V_\text{BD}^\text{init}$. This method provides a robust estimation of $V_\text{BD}^\text{init}$ consistent with the onset of avalanche pulses. The initial value of the critical voltage ($V_\text{cr}$) parameter is considered as the maximum value of the voltage. The initial value of the parameter \changed{$A$} is simply set as the maximum current value. To find $I_\text{leak}$ parameters \changed{$a$} and \changed{$b$}, the current--voltage data were first restricted to the pre-breakdown region by selecting points with $V < V_\text{BD}^\text{init}$, where the current exhibits an approximately exponential dependence on bias voltage. From the \changed{pre-breakdown subset of the reverse I--V data}, two consecutive data points closest to the midpoint of the voltage range were selected to represent the central region of interest.

The initial values of the parameters \changed{$a$ and $b$}  were extracted by linearizing the equation \ref{eq:leak} by taking the log of both sides as shown below:
\begin{equation}
\ln I = aV + b,
\end{equation}
solving analytically using the two selected points. The slope $a$ and the intercept $b$ were obtained as
\begin{equation}
a = \frac{\ln I_2 - \ln I_1}{V_2 - V_1}, \qquad
b = \ln I_1 - aV_1.
\end{equation}
This approach provides a fast estimate of the local exponential behavior of the current in the pre-breakdown region.
\par
\begin{figure}[H]
    
    \centering{\includegraphics[width=0.6\textwidth]{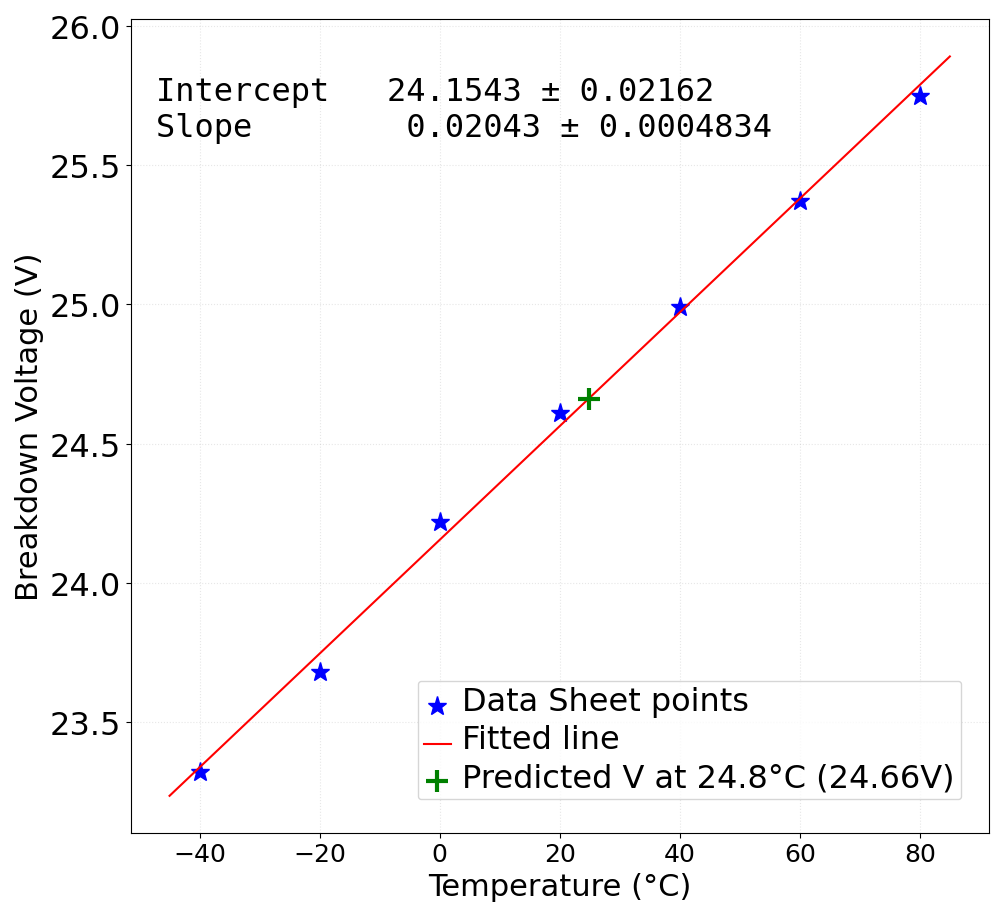}}
    
    \caption{\label{fig:sipm_temp_curr_vbd}Linear fit to the breakdown voltage versus temperature data from Table~\ref{tab:sipm_temp_data}.}
    
\end{figure}

 The fit parameters obtained from the I--V data are shown in figure ~\ref{fig:break_graph_fit}. The extracted breakdown voltage at $24.8~^\circ\mathrm{C}$ is $24.65~\mathrm{V}$. To verify the results, we digitized the data points from the I-V plot in the manufacturer's datasheet \cite{onsemi_sipm_datasheet_2022} for the same SiPM model ( MicroFC-60035-SMT) using the Plot Digitiser application \cite{plotdigitizer2026}, as shown in figure \ref{fig:data_sheet}. The digitized breakdown voltages ($V_\text{BD}$) and the current value at 2.5 V over-voltage and at different temperatures are tabulated in table \ref{tab:sipm_temp_data}. \changed{The extracted data were then fitted with a linear function, yielding a breakdown voltage of 24.66 V at 24.8 $^\circ\mathrm{C}$, which  is in very good agreement with our results as shown in figure figure~\ref{fig:sipm_temp_curr_vbd}. The breakdown voltage increases linearly with temperature at a rate of approximately 20.4~mV/$^\circ$C, as shown in figure~\ref{fig:sipm_temp_curr_vbd}, consistent with typical SiPM behaviour.} 
The Geiger probability $P_{\mathrm{Geiger}}(V)$ versus bias voltage is shown in the lower panel of figure~\ref{fig:break_graph_fit}, using a fitted $p$ value of 0.346 and a breakdown voltage $V_\text{BD}$ of $24.7\,\mathrm{V}$.
 
The total dark current ($I_{\mathrm{dark}}$) at an overvoltage of 2.5~V, corresponding to a bias voltage of 27.15~V, is found to be 1519.49~nA, as shown in figure.~\ref{fig:break_graph_fit}. According to the manufacturer's datasheet~\cite{onsemi_sipm_datasheet_2022} for the same SiPM model (MicroFC-60035-SMT), the dark current at 21~$^\circ\mathrm{C}$ ranges from 618-1750~nA. However, the dark current value at 24.8~$^\circ\mathrm{C}$ is not provided in the datasheet \cite{onsemi_sipm_datasheet_2022}.

\begin{figure}[H]
    \centering{\includegraphics[width=0.7\textwidth]{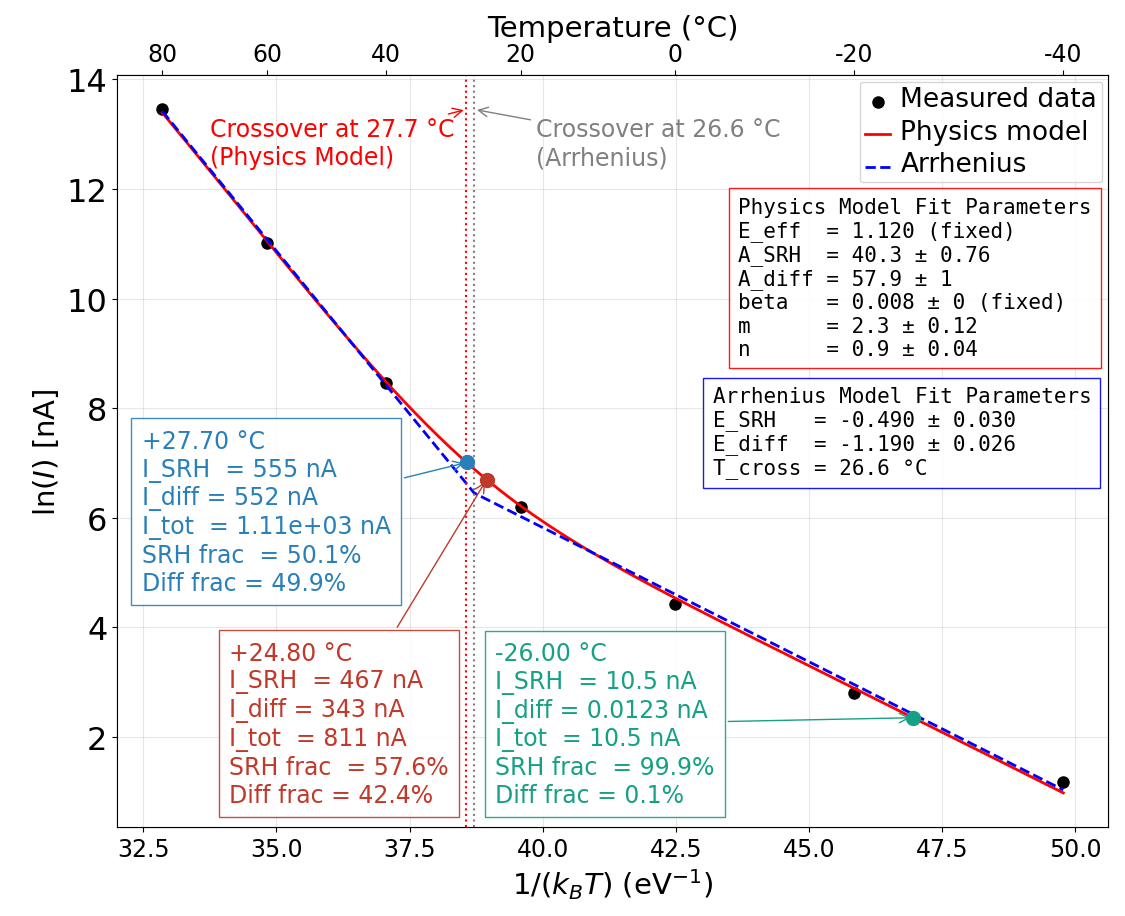}}
    \caption{\label{fig:sipm_temp_curr}$\ln(I_{dark})$ vs $1/k_B T$ data from Table \ref{tab:sipm_temp_data} fitted with equations \ref{eqn:I_dark_full} and \ref{eqn:Arr}, showing the SRH- and diffusion-dominated regions of the readymade module's dark current (Arrhenius analysis).}
    
\end{figure}

Since the dark current is temperature dependent, as discussed in Section~\ref{sec:temp_dark}, its value at the reference temperature of 24.8~$^\circ\mathrm{C}$ can be estimated as 811~nA by fitting $\log\!\left(I_{\mathrm{dark}}^{\mathrm{datasheet}}\right)$ versus $1/(k_B T)$ using the data in table~\ref{tab:sipm_temp_data} with equation~\ref{eqn:I_dark_full}, as shown in figure~\ref{fig:sipm_temp_curr}. We refer to this approach as the \emph{Physics Model}. Since the Si bandgap  energy $E_{\mathrm{eff}}$ has a weak dependence on temperature~\cite{VARSHNI1967149}, it is fixed at 1.12~eV in the fit. The gain parameter $\beta$ is fixed at 0.008~$\mathrm{K^{-1}}$ \cite{onsemi_sipm_datasheet_2022}. The parameters $m$ and $n$ are tightly constrained around 2 and 1, respectively, while the remaining parameters are allowed to float. 

Although the estimated value of 811~nA differs from the measured value of 1519~nA by 708~nA, this discrepancy is acceptable, since the dark current, even for the same SiPM model operated at fixed overvoltage and temperature, can vary significantly due to intrinsic device-to-device variations and differences in defect density, afterpulsing, and optical crosstalk, as also evident from the manufacturer's datasheet \cite{onsemi_sipm_datasheet_2022}.

It is interesting to note that at 24.80~$^\circ\mathrm{C}$, the contribution of the current from SRH generation ($I_{\mathrm{SRH}}$) is 57.6\%, while that from diffusion ($I_{\mathrm{Diff}}$) is 42.4\%, as explicitly labeled in figure~\ref{fig:sipm_temp_curr}. Since, in this work, only two generation mechanisms---SRH and diffusion---are considered, the contributions from SRH and diffusion are expected to be equal (50\% each) at the crossover temperature, which is at 27.7~$^\circ\mathrm{C}$ (also labeled in the figure, at 27.7~$^\circ$C: 50.1\%/49.9\%). On this basis, the SRH-dominated and diffusion-dominated regions can be clearly separated.

To cross-check the model, we evaluated the current contributions at $-26~^\circ\mathrm{C}$, where the SRH fraction is 99.9\% and the diffusion fraction is 0.1\%, which is consistent with theoretical expectations (see figure~\ref{fig:sipm_temp_curr}). The fitted values of the parameters $m$ and $n$ are 2.3 and 0.9, respectively, which are also consistent with the theory discussed earlier in section~\ref{sec:temp_dark}. However, the relatively high value of $m = 2.3$ suggests that, for accurate modeling of dark-current   additional process, such as electric-field-induced contributions, should be included in the dark current modelling, as described in~\cite{Engelmann_2016}. 

\changed{In a semiconductor, the generation of a free carrier that contributes to
the dark current requires thermal excitation across an energy barrier.
This excitation can occur either directly across the full bandgap, via
band-to-band (or diffusion-mediated) generation, or via an intermediate
defect level within the bandgap, i.e., Shockley--Read--Hall (SRH)
generation, in which case the effective barrier is reduced to
approximately half the bandgap for a mid-gap trap. The activation
energy $E_a$, obtained from the slope of $\ln(I)$ versus $1/k_BT$ in an
Arrhenius plot, is therefore a direct measurement of the height of this
thermal barrier for whichever generation mechanism dominates the dark
current at a given temperature. Consequently,  the  magnitude of $E_a$ identifies
the microscopic origin of the dominant dark-current generation process,
distinguishing bulk diffusion ($E_a \approx E_g$) from defect-mediated
SRH generation ($E_a \approx E_g/2$). A SiPM exhibiting anomalous activation energies in either regime may signal elevated defect density or radiation damage, making the Arrhenius slope a rapid screening diagnostic during batch qualification campaigns.} As suggested in~\cite{pagano}, the activation energy can be determined by neglecting the $T^{3/n}$ term in equation~\ref{eq:arr_eq}, since it has a weak dependence on the dark current $I_{\mathrm{SRH/Diff}}$ \cite{Sze_Ng_Physics_Semiconductor_Devices}. Thus, by retaining only the exponential (Arrhenius) term and linearising the expression, the slope yields the activation energy in different temperature regions as shown below:
\begin{equation} \label{eqn:Arr}
\ln(I) = -\frac{E_a}{k_B\,T} + \ln(A_{SRH/Diff})
\end{equation}
 
We use the Python \texttt{PWLF} \cite{pwlf} library to fit the data using equation~\ref{eqn:Arr}, as it automatically estimates the crossover point (26.6~$^\circ\mathrm{C}$) and fits two linear regions using the same estimated crossover point, as shown in figure~\ref{fig:sipm_temp_curr} (Arrhenius model). The observed activation energy at higher temperatures is close to the silicon band-gap energy $E_g$, which is 1.19~eV (ideally 1.12~eV at room temperature). The activation energy at low temperatures is found to be 0.49~eV, which is approximately $E_a \approx E_g/2$, implying the presence of traps with ionization energies close to the silicon mid-gap. These defects act as efficient Shockley--Read--Hall (SRH) generation centers and dominate the leakage current in the low-temperature regime.

\begin{table}[h!]
    \centering
    \caption{Temperature dependence of breakdown voltage and dark current. \changed{Values of $V_\text{BD}$ and $I_\text{dark}^\text{datasheet}$ were obtained by digitizing the I--V curves from the manufacturer's datasheet~\cite{onsemi_sipm_datasheet_2022} using the PlotDigitizer application~\cite{plotdigitizer2026}.}}
    \label{tab:sipm_temp_data}
    \begin{tabular}{ccccc}
        \toprule
        Temp ($^\circ$C) & $1/(k_B T)$ ($\text{eV}^{-1}$) & $V_{BD}$ (V) & $V_{BD} + 2.5$ (V) & $I_{dark}^{data\,sheet}$ (nA) \\
        \midrule
        -40 & 49.79 & 23.32 & 25.83 & 3.25 \\
        -20 & 45.85 & 23.68 & 26.18 & 16.58 \\
        0   & 42.50 & 24.22 & 26.72 & 84.49 \\
        20  & 39.58 & 24.61 & 27.11 & 495.59 \\
        40  & 37.03 & 24.99 & 27.49 & $4.69 \times 10^{3}$ \\
        60  & 34.80 & 25.37 & 27.88 & $6.03 \times 10^{4}$ \\
        80  & 32.84 & 25.75 & 28.25 & $6.94 \times 10^{5}$ \\
        \bottomrule
    \end{tabular}
\end{table}

\changed{\subsubsection{Results of Modules Fabricated at SINP}}\label{sec:sinp_results}

\begin{figure}[httb]
    \centering{\includegraphics[width=1.\textwidth]{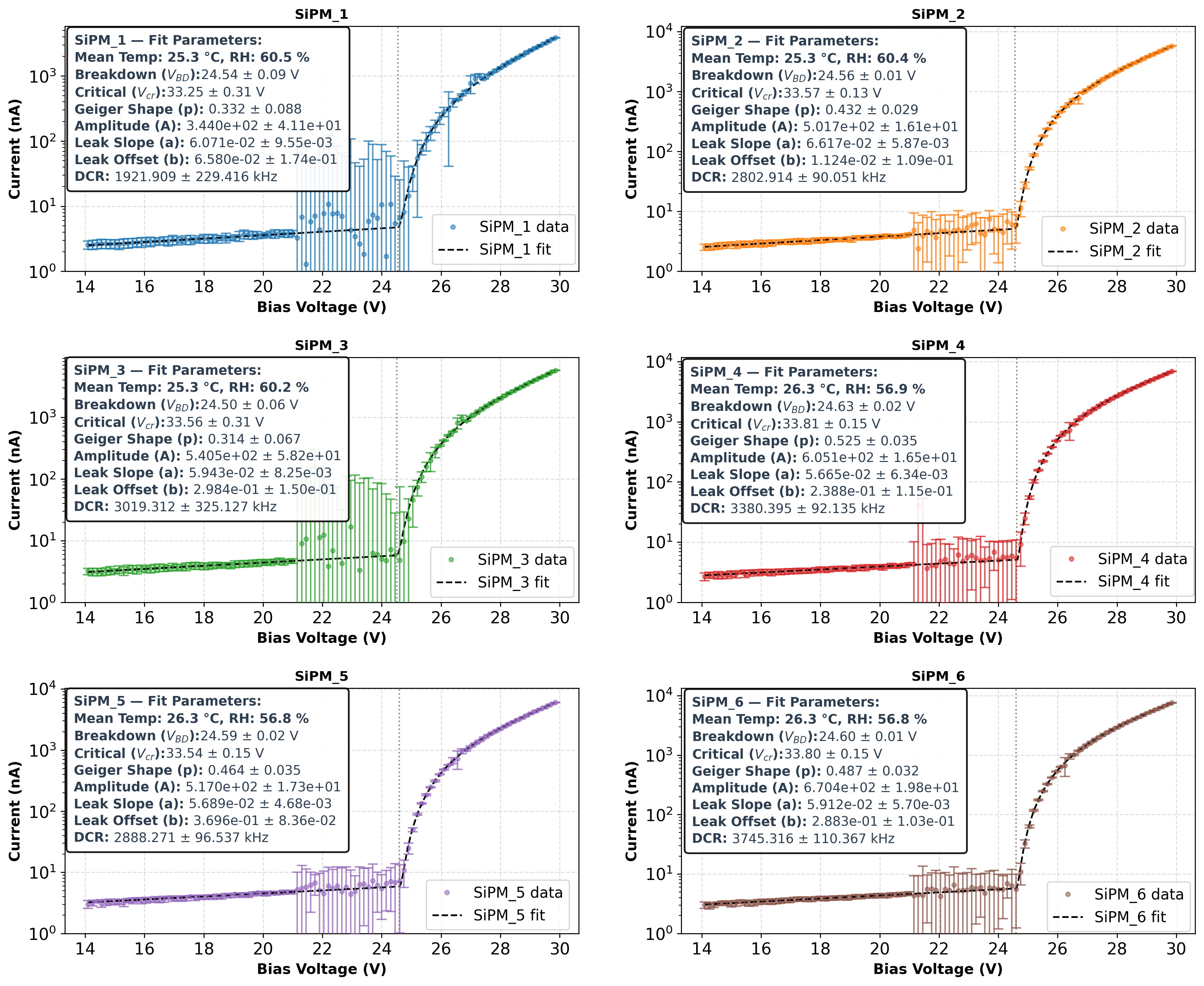}}
    \caption{\label{fig:batch_results}\changed{Reverse-bias $I$--$V$ characterization of six SiPM modules fabricated and packaged at SINP (SiPM\_1--SiPM\_6), measured simultaneously in batch mode using the multi-channel bias distribution board of Section~\ref{sec:bias_distribution}. Points show the measured dark current with statistical error bars; dashed lines show the fit to equation~\ref{eq:idark}. Fitted parameters -- mean temperature and relative humidity, breakdown voltage, critical voltage, Geiger shape parameter, amplitude, leakage-current parameters, and derived DCR -- are listed in each panel.}}
    
\end{figure}

\changed{
To validate the multi-channel bias distribution board (section~\ref{sec:bias_distribution}), six custom SiPM modules were tested simultaneously. These modules (SiPM\_1--SiPM\_6) were fabricated at SINP and distributed across two boards, each equipped with three channels (see figure~\ref{fig:bias_distribution}). The automated batch $I$--$V$ sequencer from section~\ref{sec:iv_logic} managed the characterization. It swept each channel over the exact same reverse-bias voltage range. Afterward, each dataset was fitted independently to equation~\ref{eq:idark}. This followed the same initialization procedure used for the readymade module in section~\ref{sec:results_readymade}. The resulting fits are shown in figure~\ref{fig:batch_results}, with the extracted parameters listed in each panel.
 
 The mean breakdown voltage of the six SiPMs is $V_\text{BD} = 24.57\,\mathrm{V}$ with a standard deviation of $0.04\,\mathrm{V}$. The difference between the highest (24.63~V) and lowest (24.50~V) measured values is a narrow range of only 0.13~V, which demonstrates a high degree of consistency across the devices. These results align with the manufacturer's datasheet, which specifies 24.2--24.7~V at 21~$^\circ$C (Section~\ref{sec:detector_description}). To verify this, we applied a temperature correction using the 20.4~mV/$^\circ$C coefficient from figure~\ref{fig:sipm_temp_curr_vbd}. At our recorded operating temperature of $\sim$25.3--26.3~$^\circ$C, the projected datasheet range shifts to approximately 24.3--24.8~V. Our observed values fall comfortably within this adjusted range. This strong agreement confirms both the reliability of the automated fitting procedure and the uniform fabrication quality of the SINP devices. Finally, it demonstrates that the relay-switched multi-channel board introduces no measurable systematic bias compared to the single-channel measurements in section~\ref{sec:results_readymade}.

By contrast, the fitted Geiger shape parameter $p$ shows substantially more scatter (0.314--0.525) than $V_\text{BD}$, and the pre-breakdown leakage parameters $a$ and $b$ vary by close to an order of magnitude between channels. For all SiPMs, the leakage current becomes highly unstable near the breakdown voltage, resulting in large errors for parameters $a$ and $b$ in this region.

The fitted amplitude $A$, and the dark count rate derived from it, are discussed together with the readymade-module reference value in section~\ref{sec:dcr}.
}

\changed{The fitted parameters for all six channels are summarized in table~\ref{tab:sinp_batch_params}, together with the total current $I_\text{tot} = I_\text{leak} + I_\text{aval}$ evaluated at an overvoltage of 2.5~V (i.e.\ at $V = V_\text{BD}+2.5$~V) using equation~\ref{eq:idark}, for direct comparison with the readymade-module result of Section~\ref{sec:results_readymade}. The total current at 2.5~V overvoltage ranges from 685.8~nA (SiPM\_1) to 1626.9~nA (SiPM\_6).
The 1519.49~nA obtained for the readymade evaluation module at 24.8~$^\circ$C (Section~\ref{sec:results_readymade}) falls within one standard deviation of the batch mean, $I_\text{tot}(2.5\,\text{V OV}) = 1208.5 \pm 313\,\mathrm{nA}$. This indicates that the readymade module is not an outlier relative to the spread present among the SINP-fabricated devices.
}

\begin{figure}[H]
    \centering
    \includegraphics[width=0.9\textwidth]{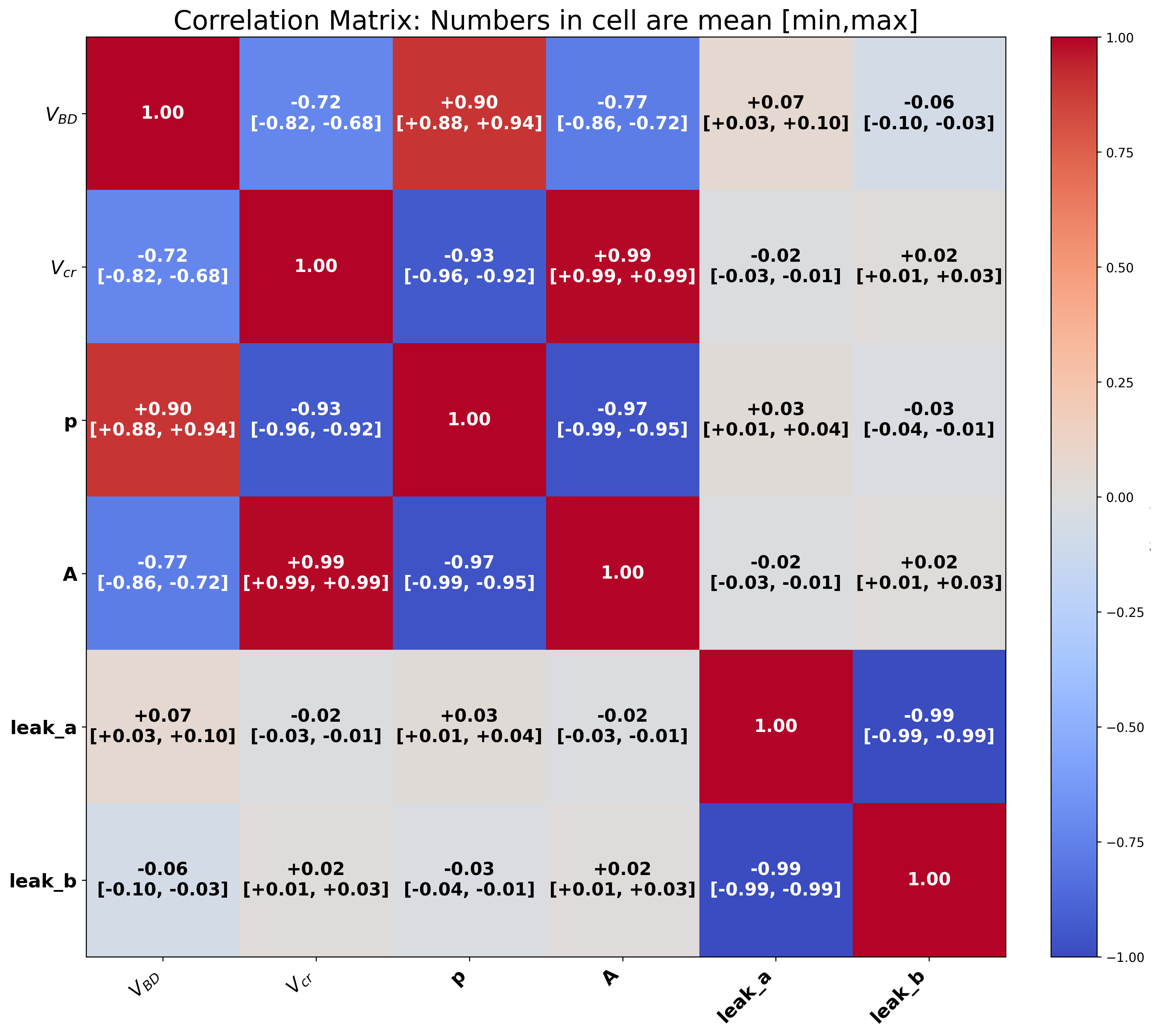}
    \caption{\changed{Mean Pearson correlation between reverse-bias fit parameters (equation~\ref{eq:idark}), averaged over the six SINP-fabricated SiPM channels of figure~\ref{fig:batch_results}. Each cell reports the mean correlation across channels, with the [minimum, maximum] range observed across channels given below it.}}
    \label{fig:corr_summary}
\end{figure}

\begin{table}[h!]
\centering
\caption{\changed{Fitted $I$--$V$ parameters for the readymade MICROFC-SMA-60035-GEVB module (figure~\ref{fig:break_graph_fit}) and the six SINP-fabricated SiPM modules (figure~\ref{fig:batch_results}), together with the leakage current ($I_\text{leak}$), avalanche current ($I_\text{aval}$), and total current ($I_\text{tot}$) evaluated at 2.5~V overvoltage using equation~\ref{eq:idark}.}}
\label{tab:sinp_batch_params}
\resizebox{\textwidth}{!}{%
\begin{tabular}{lccccccc}
\toprule
Parameter & \shortstack{MICROFC-SMA\\-60035-GEVB} & SiPM\_1 & SiPM\_2 & SiPM\_3 & SiPM\_4 & SiPM\_5 & SiPM\_6 \\
\midrule
Mean Temp ($^\circ$C) & 24.8 & 25.3 & 25.3 & 25.3 & 26.3 & 26.3 & 26.3 \\
RH (\%) & 59.7 & 60.5 & 60.4 & 60.2 & 56.9 & 56.8 & 56.8 \\
$V_\text{BD}$ (V) & 24.65 & 24.54 & 24.56 & 24.50 & 24.63 & 24.59 & 24.60 \\
$V_\text{cr}$ (V) & 33.47 & 33.25 & 33.57 & 33.56 & 33.81 & 33.54 & 33.80 \\
Geiger shape $p$ & 0.346 & 0.332 & 0.432 & 0.314 & 0.525 & 0.464 & 0.487 \\
Amplitude $A$ (nA/V) & $7.496\times10^{2}$ & $3.440\times10^{2}$ & $5.017\times10^{2}$ & $5.405\times10^{2}$ & $6.051\times10^{2}$ & $5.170\times10^{2}$ & $6.704\times10^{2}$ \\
Leak slope $a$ & $6.243\times10^{-2}$ & $6.071\times10^{-2}$ & $6.617\times10^{-2}$ & $5.943\times10^{-2}$ & $5.665\times10^{-2}$ & $5.689\times10^{-2}$ & $5.912\times10^{-2}$ \\
Leak offset $b$ & $-2.204\times10^{-1}$ & $6.580\times10^{-2}$ & $1.124\times10^{-2}$ & $2.984\times10^{-1}$ & $2.388\times10^{-1}$ & $3.696\times10^{-1}$ & $2.883\times10^{-1}$ \\
DCR (kHz) & 4187.6 & 1921.9 & 2802.9 & 3019.3 & 3380.4 & 2888.3 & 3745.3 \\
$I_\text{leak}$ at $V_\text{BD}+2.5$~V (nA) & 4.37 & 5.51 & 6.06 & 6.71 & 5.90 & 6.76 & 6.62 \\
$I_\text{aval}$ at $V_\text{BD}+2.5$~V (nA) & 1515.12 & 680.2 & 1146.4 & 1015.0 & 1519.4 & 1231.2 & 1620.2 \\
$I_\text{tot}$ at $V_\text{BD}+2.5$~V (nA) & \textbf{1519.49} & \textbf{685.8} & \textbf{1152.5} & \textbf{1021.7} & \textbf{1525.3} & \textbf{1238.0} & \textbf{1626.9} \\
\bottomrule
\end{tabular}}
\end{table}
\changed{
\subsubsection{Parameter Correlations Across the Batch}
The total measured dark current ($I_{tot}$) is modeled as the sum of two distinct physical processes: a pre-breakdown leakage current ($I_{leak}$) and a post-breakdown avalanche current ($I_{aval}$). As demonstrated by the I-V characteristics across all tested channels (see figure~\ref{fig:batch_results}), the model accurately captures the transition between the flat pre-breakdown baseline and the steep avalanche multiplication. 

To evaluate the robustness of these fits, the parameter correlation matrix was analyzed, as shown in figure~\ref{fig:corr_summary}. The leakage parameters ($a$ and $b$) are entirely uncorrelated with the avalanche parameters ($|\rho| \lesssim 0.1$), confirming that the fitting algorithm successfully isolates the initial leakage baseline without being skewed by the subsequent avalanche rise. However, $a$ and $b$ exhibit a strong anti-correlation with each other ($\rho \approx -0.99$), which is the standard mathematical expectation for the slope and intercept of an exponential baseline. 

Within the avalanche regime, the parameters ($V_{BD}$, $V_{cr}$, $p$, $A$) exhibit expected correlations as they jointly govern the shape of the post-breakdown curve. Most notably, the critical voltage ($V_{cr}$) and the amplitude factor ($A$) demonstrate a near-degeneracy ($\rho \approx +0.99$). Because the upper limit of the measurement sweep ($30$~V) remains below the extracted critical voltage ($V_{cr} \approx 33.5$~V), the data does not capture the true divergence of the afterpulsing factor. Consequently, the fit can compensate for an increase in the divergence pole ($V_{cr}$) by adjusting the amplitude ($A$) without reducing the overall fit quality. Importantly, this inherent trade-off does not compromise the stability of the fit or the precision of the extracted breakdown voltage ($V_{BD}$).}
\subsubsection{Estimation of Dark Count Rate (DCR)}\label{sec:dcr}
The DCR can be estimated from the $I$--$V$ characteristics for quick verification. In the dark current equation~\ref{eq:idark}, the term $\frac{dN_{\text{carriers}}}{dt}$ explicitly represents the rate of free carriers generated in the bulk silicon that initiate Geiger discharges, which is equivalent to the DCR.

From the fitted value of the parameter $A$, the DCR can be estimated if the value of $C_{\mu\text{cell}}$ is known (typically available in the datasheet), as given by
\begin{equation}\label{eqn:DCR}
    A \approx \text{DCR} \cdot C_{\mu\text{cell}} .
\end{equation}

The fitted value of $A$ is $7.496 \times 10^{2}$~nA/V, and in our case $C_{\mu\text{cell}}$ is $1.79 \times 10^{-13}$~F according to the datasheet \cite{onsemi_sipm_datasheet_2022}. Substituting these values in equation \ref{eqn:DCR}, the DCR is estimated to be approximately 4.19~MHz at our reference temperature $24.8~^\circ\mathrm{C}$ . As discussed in section~\ref{sec:detector_description}, the DCR at $21~^\circ\mathrm{C}$ lies in the range 1.2--3.4~MHz; therefore, a higher value is expected at elevated temperatures. This approach provides a rapid method for qualifying the noise performance of the SiPM using only dark current measurements. However, for more precise measurement one need to perform waveform analysis of SiPM pulses.

\par\changed{The same method was applied to the six SINP-fabricated modules characterized in batch mode in Section~\ref{sec:sinp_results} (figure~\ref{fig:batch_results}). The resulting per-channel DCR values range from 1.92 to 3.75~MHz, with a batch mean of $2.96 \pm 0.56$~MHz (mean $\pm$ standard deviation), computed using the same $C_{\mu\text{cell}}$ as the readymade module. This is comparable to, though somewhat lower than, the 4.19~MHz obtained for the readymade evaluation module at 24.8~$^\circ$C; the difference is consistent with the device-to-device variation in defect density already noted in Section~\ref{sec:results_readymade}, and may also reflect genuinely lower bulk-defect DCR in the SINP-fabricated devices.
}
\changed{
\subsection{Determination of Quenching Resistance}\label{sec:quench}

\subsubsection{Physical Origin of the Quenching Resistance}
Each SiPM microcell consists of an APD operated above breakdown in series with an integrated
polysilicon resistor, $R_q$ ~\cite{buzhan2003, mazzillo2008}. When a photon or a thermally generated carrier triggers a Geiger-mode
avalanche, the resulting current pulse flows through $R_q$, producing a voltage drop that reduces
the effective bias across the microcell junction below $V_\text{BD}$. This passively quenches the
avalanche without requiring any active feedback circuit. Once quenched, the microcell recharges
through the same resistor with a characteristic time constant $\tau_\text{recovery} \approx R_q
\cdot C_\text{cell}$, where $C_\text{cell}$ is the microcell capacitance. Since $R_q$ is a polysilicon structure defined during fabrication, its value and uniformity across microcells and across devices serve as a sensitive probe of wafer-level process quality.

Since all $N_\text{microcell}$ resistors are nominally identical and connected in parallel when the
device is forward-biased as a whole, the resistance seen at the external terminals in the ohmic
(linear) region of the forward $I$--$V$ curve is
\begin{align}
R_\text{total} &= R_\text{bias} + \frac{R_q}{N_\text{microcell}}, \\
R_q &= N_\text{microcell}\times(R_\text{total}-R_\text{bias}) \label{eq:rq_total}
\end{align}
where $R_\text{bias}$ is the resistance external to the sensor (bias-circuit series resistors and
readout load, figure~\ref{fig:sipm_bias}). $R_\text{total}$ is obtained from the inverse slope of a
linear fit, $I = mV + c$, to the ohmic knee region of the forward-bias curve, so that
$R_\text{total} = 1/m$. It should be noted that when determining the total quenching resistance, any supplementary series resistance present along with the SiPM bias circuit must be added to the SiPM bias resistance ($R_{\mathrm{bias}}$). A more detailed discussion on the quenching resistance can be found in \cite{Acerbi_2026}.
 \begin{figure}[htbp]
    \centering{\includegraphics[width=0.7\textwidth]{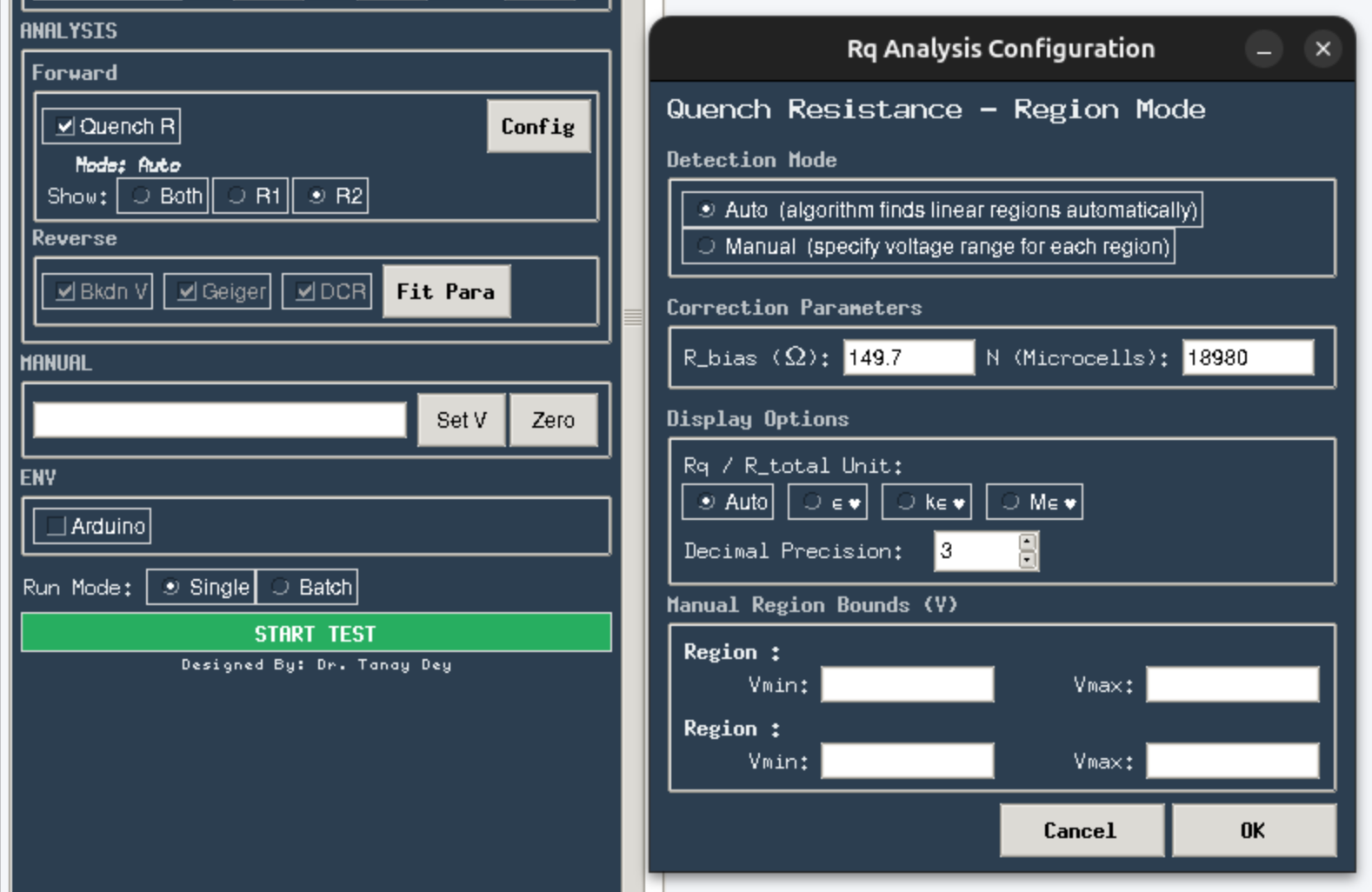}}
    \caption{ \label{fig:rq_config}\changed{Settings panel for quenching-resistance extraction, featuring the CONFIG tab to define required parameters such as the bias resistance ($R_\text{bias}$), microcell count ($N_\text{microcell}$), and the fitting range.}
    \label{fig:rq_config}}
\end{figure}
\subsubsection{Quenching-Resistance Analysis using PySiPMGUI}

The forward-bias measurement is configured similarly to the reverse-bias sweep described in Section \ref{subsec:acquisition}, requiring only that the polarity toggle be set to \texttt{Forward} (figure \ref{fig:gui_screenshot}). During or after data acquisition, the quenching resistance ($R_q$) analysis can be overlaid on the real-time plot. The parameters used to isolate the linear region and compute $R_q$ are managed through a dedicated configuration dialog (figure \ref{fig:rq_config}), which provides two distinct detection modes. By default, the user can select the automated detection option, allowing the software to autonomously identify the linear ohmic region of the forward $I-V$ curve.  Alternatively, the user can choose manual detection to explicitly define the linear range by setting the exact voltage boundaries (\texttt{Vmin} and \texttt{Vmax}). 
The \texttt{Correction Parameters} fields in the same dialog set $R_\text{bias}$ (in $\Omega$) and
$N_{\text{microcell}}$, which are substituted into equation~\ref{eq:rq_total} to convert the fitted slope
into $R_q$; these can be overridden on a per-channel basis during batch runs (per-channel
$R_\text{bias}/N$ fields in the batch configuration panel as shown in figure \ref{fig:batch_config}), or
entered in the main forward-bias panel as shown in figure \ref{fig:rq_config}. The linear fit itself is performed as a $\chi^2$
minimization of $I = mV + c$ against the measured current, weighted by the per-point standard
deviation over the $N$ repeated measurements at each voltage step (Section~\ref{subsec:acquisition})
when available; the fitted slope uncertainty $\sigma_m$ is propagated to $R_q$ via $\sigma_{R_q} =
N_\text{microcell}\cdot\sigma_m/m^2$.

\subsubsection{Results of Quenching Resistance of the MICROFC-SMA-60035-GEVB Module and the SINP-Fabricated Batch}
\changed{The forward-bias I--V characteristic curves for the MICROFC-SMA-60035-GEVB module and the six SINP-fabricated modules are shown in figure~\ref{fig:readymade_quenching_results} and figure~\ref{fig:quenching_results}, respectively. In each case, the linear fit range is set to 1.30--1.60~V. The slope of the fit, $m$, yields the total series resistance, $R_\text{total} = 1/m$. By combining this value with the external bias-circuit resistance, $R_\text{bias}$, and the microcell count, $N_\text{microcell} = 18980$, equation~\ref{eq:rq_total} gives a quenching resistance of $R_q = 546.622 \pm 0.020\,\mathrm{k\Omega}$ for the readymade module. The corresponding values for the six SINP channels, derived using the identical fit window and analysis procedure, are summarized alongside this reference measurement in table~\ref{tab:sinp_quenching_params}
}
\begin{figure}[H]
    \centering
    \subfloat[\label{fig:readymade_quenching_results}]{\includegraphics[width=.55\textwidth]{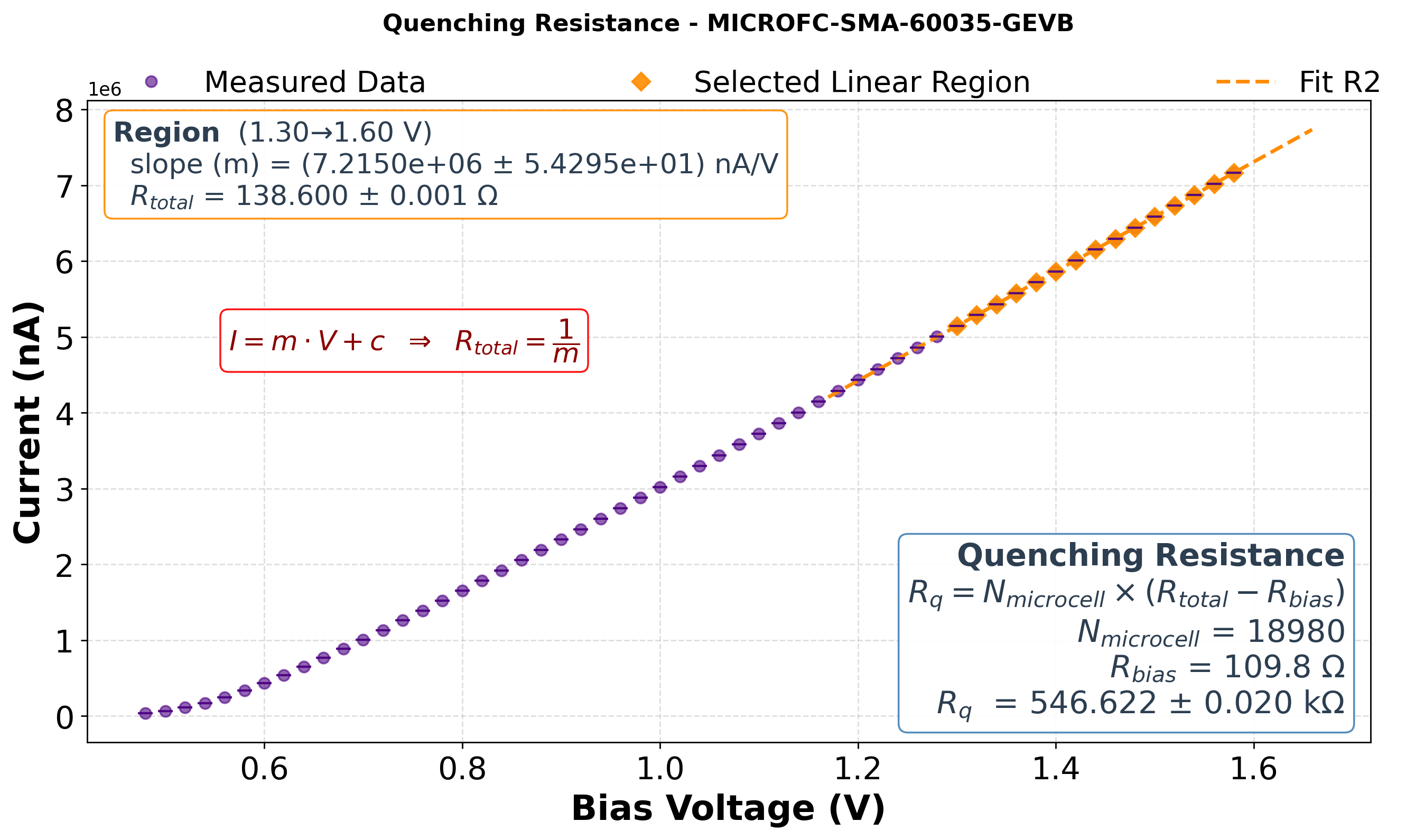}}\\
    \subfloat[\label{fig:quenching_results}]{\includegraphics[width=1.\textwidth]{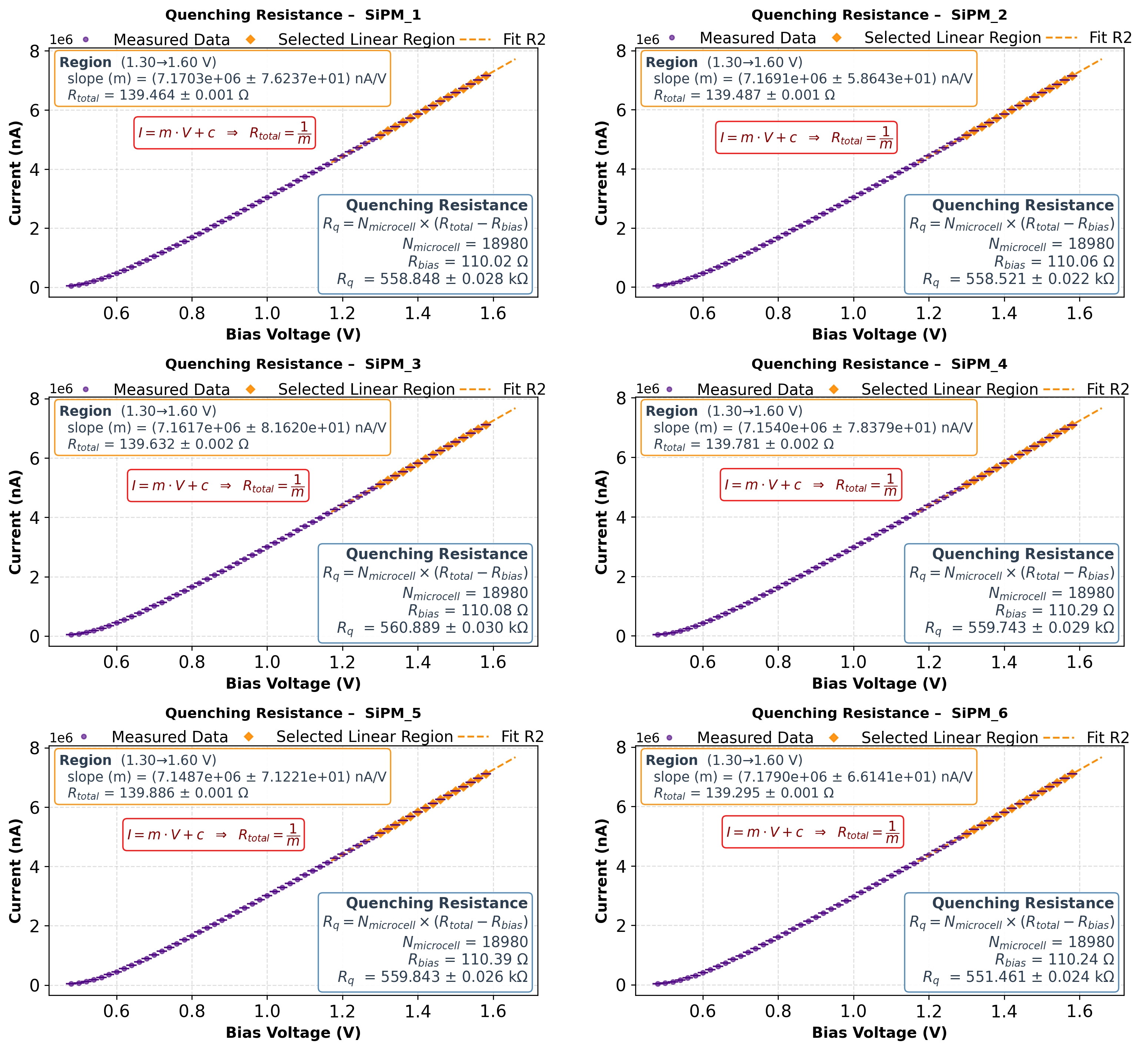}}
    \caption{\changed{Forward-bias $I$--$V$ characterization of the (a) MICROFC-SMA-60035-GEVB and (b) six SINP-fabricated SiPM modules, used to extract the quenching resistance. Purple points show the measured forward current; orange points mark the selected ohmic region (1.30--1.60~V) used for the fit.}}
\end{figure}
\begin{table}[h!]
    \centering
    \caption{\changed{Fitted quenching-resistance parameters for the readymade MICROFC-SMA-60035-GEVB module (figure~\ref{fig:readymade_quenching_results}) and the six SINP-fabricated SiPM modules (figure~\ref{fig:quenching_results}). The linear fit $I = mV + c$ is performed over the linear region from 1.30~V to 1.60~V of the forward-bias $I$--$V$ curve for each module.}}
    \label{tab:sinp_quenching_params}
    \resizebox{\textwidth}{!}{%
    \begin{tabular}{lccccccc}
        \toprule
        Parameter & \shortstack{MICROFC-SMA\\-60035-GEVB} & SiPM\_1 & SiPM\_2 & SiPM\_3 & SiPM\_4 & SiPM\_5 & SiPM\_6 \\
        \midrule
        Fit region (V)                                   & \multicolumn{7}{c}{1.30 -- 1.60 (all channels/module)} \\
        Slope $m$ (nA/V)                                 & $7.2150\times10^{6}$ & $7.1703\times10^{6}$ & $7.1691\times10^{6}$ & $7.1617\times10^{6}$ & $7.1540\times10^{6}$ & $7.1487\times10^{6}$ & $7.1790\times10^{6}$ \\
        $R_\text{total} = 1/m$ ($\Omega$)                & 138.600 & 139.464 & 139.487 & 139.632 & 139.781 & 139.886 & 139.295 \\
        $R_\text{bias}$ ($\Omega$)                       & 109.8   & 110.02  & 110.06  & 110.08  & 110.29  & 110.39  & 110.24  \\
        $R_\text{total} - R_\text{bias}$ ($\Omega$)      & 28.800  & 29.444  & 29.427  & 29.552  & 29.491  & 29.496  & 29.055  \\
        $N_\text{microcell}$                             & \multicolumn{7}{c}{18980 (all channels/module)} \\
        $R_q$ (k$\Omega$)                                & \textbf{546.622} & \textbf{558.848} & \textbf{558.521} & \textbf{560.889} & \textbf{559.743} & \textbf{559.843} & \textbf{551.461} \\
        \bottomrule
    \end{tabular}}
\end{table}
Table~\ref{tab:sinp_quenching_params} summarizes the fitted linear-region parameters and the derived quenching resistance for all six channels, together with the readymade MICROFC-SMA-60035-GEVB module. The extracted mean and standard deviation quenching resistance for the SINP batch is $R_q = 558.2 \pm 3.1~\mathrm{k\Omega}$, with a maximum spread of only 9.4~k$\Omega$ across the six channels, corresponding to less than 2\% of the mean value. This is the tightest clustering observed among all fitted parameters for this batch. The microcell quenching resistor is formed from a thin-film polysilicon layer integrated during SiPM fabrication~\cite{buzhan2003, mazzillo2008}, so this level of channel-to-channel uniformity reflects good wafer-level process consistency across the six independently packaged devices.

The readymade module's quenching resistance, $R_q = 546.622 \pm 0.020~\mathrm{k\Omega}$, is about 2.1\% below the SINP batch mean and just outside its $\pm 3.1~\mathrm{k\Omega}$ spread. Since $R_q$ depends on the polysilicon sheet resistance and geometry set during fabrication, this small offset is consistent with normal batch-to-batch process variation between the readymade unit and the SINP-fabricated devices, rather than indicating a measurement issue.
}
\section{Summary \& Discussion }
\label{sec:summary} 
We have developed an open-source, Python-based graphical user interface (GUI) for characterizing Silicon Photomultipliers, offering a flexible and cost-effective alternative to proprietary measurement software. Built on the PyVISA framework, the software provides automatic control of commonly used laboratory power supplies and enables direct extraction of key parameters from current-voltage measurements. To ensure device safety, we prioritized controlled voltage ramping and continuous current monitoring, which are essential for large-scale testing.
 
Breakdown voltage is determined by fitting the full I--V curve with a physics-based model, avoiding derivative-based methods that are sensitive to noise. This model also enables estimation of the primary dark count rate from DC measurements, providing a fast and reliable tool for sensor qualification. \changed{To validate the analysis procedure, results were compared against the manufacturer's specifications for the SensL MicroFC-60035-SMT \cite{onsemi_sipm_datasheet_2022}. The fit yielded a breakdown voltage of 24.65~V at 24.8~$^\circ$C, in excellent agreement with the digitized reference value of 24.66~V, alongside a DCR estimate of 4.19~MHz that is consistent with expected temperature scaling. The same procedure, applied in batch mode to six SINP-fabricated modules, returned breakdown voltages ranging from 24.50 to 24.63~V at temperatures between 25.3 and 26.3~$^\circ$C, alongside DCR values of 1.92--3.75~MHz. This confirms that the multi-channel bias-distribution board introduces no measurable systematic bias relative to the single-channel measurement. The forward-bias quenching-resistance analysis yielded $R_q = 546.6\,\mathrm{k\Omega}$ for the readymade module and $558.2 \pm 3.1\,\mathrm{k\Omega}$ for the SINP-fabricated batch, indicating good wafer-level process uniformity across the tested devices.}
 
The dependence of the dark current on the temperature is discussed. The value of dark current at an unknown temperature is estimated from the fit of the dark current vs. temperature data, which is also a quick method during batch quality assurance checks. 
\changed{To date, seven SiPM modules have been characterized using PySiPMGUI.}
 
The system is currently under development, and in the next update, we plan to add oscilloscope control features to the GUI. 

This will enable real-time characterization of the SiPM using pulse-based measurements. The parameters that can then be studied include gain (from the charge spectrum), dark count rate, optical cross-talk, and after-pulsing probability. This will extend the current DC-based analysis and make the tool more complete and robust. More generally, the modular design of the code makes it suitable for a wide range of applications, beyond the testing of SiPM arrays in astroparticle and high-energy physics. The automation, feedback, and transparent analysis support reproducible detector development.
\section{Acknowledgement}
All authors are grateful to SINP for providing infrastructure and administrative support, and to the Department of Atomic Energy (DAE) of the Government of India for providing the necessary grant. The authors thank the reviewer for constructive and meticulous review that helped improvement in both content and form of the manuscript.
\bibliographystyle{JHEP}
\bibliography{reference}

\end{document}